\documentclass[
    aps,
    pra,
    reprint,
    showpacs,
    amsmath,
    amssymb,
    amsfonts,
    lengthcheck,
    longbibliography,
    superscriptaddress
]{revtex4-2}

\usepackage{placeins}
\usepackage{graphicx}
\usepackage{dcolumn}
\usepackage{bm}
\usepackage{braket}
\usepackage{enumitem}
\usepackage[utf8]{inputenc}
\usepackage[none]{hyphenat}
\usepackage{amsmath}
\usepackage{amssymb}
\usepackage{amsthm}
\usepackage{mathtools}
\usepackage{physics}

\usepackage[
    colorlinks=true, 
    citecolor=blue,
    linkcolor=blue,
    urlcolor=blue
]{hyperref}
\usepackage{orcidlink}

\newcommand{\la}{\left\langle}
\newcommand{\ra}{\right\rangle}

\begin{document}

\title{Procrastinating einselection in non-Markovian quantum dynamics}

\author{Michael J. Moody \orcidlink{0009-0001-3065-7308}}
\affiliation{
Department of Physics,
University of Maryland, Baltimore County,
Baltimore, MD 21250, USA
}
\affiliation{
Quantum Science Institute,
University of Maryland, Baltimore County,
Baltimore, MD 21250, USA
}

\author{Tara Kalsi \orcidlink{0000-0003-1513-5795}}
\email{tkalsi@umbc.edu}
\affiliation{
Department of Physics,
University of Maryland, Baltimore County,
Baltimore, MD 21250, USA
}
\affiliation{
Quantum Science Institute,
University of Maryland, Baltimore County,
Baltimore, MD 21250, USA
}

\author{Agung Budiyono \orcidlink{0000-0002-9091-8645}}
\affiliation{
Research Center for Quantum Physics, 
National Research and Innovation Agency, 
South Tangerang 15314, Republic of Indonesia
}
\affiliation{
School of Physical and Mathematical Sciences, 
Nanyang Technological University, 
21 Nanyang Link, 637371, Singapore
}

\author{Sebastian Deffner \orcidlink{0000-0003-0504-6932}}
\affiliation{
Department of Physics,
University of Maryland, Baltimore County,
Baltimore, MD 21250, USA
}
\affiliation{
Quantum Science Institute,
University of Maryland, Baltimore County,
Baltimore, MD 21250, USA
}
\affiliation{
National Quantum Laboratory,
College Park, MD 20740, USA
}

\date{\today}

\begin{abstract}    
Among the processes governed by open-system and dissipative quantum dynamics, environment-induced superselection, or  \textit{einselection}, is particularly foundational, since it describes the formation of a preferred set of \textit{pointer states} through environmental monitoring and ultimately the emergence of classicality. Especially in the context of Quantum Darwinism, einselection can be studied in terms of the motion of an observable towards its pointer value.  To this end, we derive a quantum speed limit for observables for finite-dimensional systems governed by differentiable time-local open dynamics.  We identify two mechanisms through which non-Markovianity can enhance observable speed: direct dissipative motion supported by population retention or revival, and indirect enhancement through restoration of the asymmetry available for coherent motion. 
For permanent einselection, this delays the emergence of the pointer basis due to accumulated negativity. 
\end{abstract}

\maketitle

\section{Introduction}
\label{sec:introduction}

\emph{Quantum Darwinism} is a conceptual framework that describes the emergence of classical objectivity within our genuinely quantum Universe \cite{Zurek2025Book}. The paradigm is built upon the deep appreciation of the fact that \emph{classical objectivity} is simply the consequence of two classical observers agreeing on the physical state of a quantum object \cite{touil2025consensus}. Importantly, these observers do not measure the quantum object directly, but rather intercept only a tiny fraction of the ``environment'' that has interacted with the object. ``Darwinism'' then refers to the fact that as a consequence of decoherence only the classical information has survived, which however is encoded in the environment at enormous redundancy. 

The literature on Quantum Darwinism has often focused on thoroughly analyzing the information theoretic structure that can support redundant encoding, and hence the emergence of classical objectivity \cite{OllivierPoulinZurek2004PRL,Le2011Entropy,Cakmak2021Entropy,Le2019PRL,Ryan2021PLA,Korbicz2021quantum,Touil2022PRL,Girolami2022PRL,Acevedo2026JMP}. Decoherence, however, is a dynamical process and it appears desirable to understand not just if but also when a Universe becomes irrevocably classical \cite{JessRiedel2012NJP}. In fact, \emph{environment-induced superselection}, aka \emph{einselection} \cite{Zurek1981PRD,Zurek1982PRD,Zurek1991PT,Zurek1993PTP,Zurek2000ADP,Zurek2003RMP,Campbell2019PRA} describes how environmental interactions suppress genuine quantum correlations as quantified by quantum discord \cite{Ollivier2001PRL} and establish a preferred, or \textit{pointer} basis. Specifically, these einselected pointer states are quantum states of a system that remain stable under environmental monitoring, immune to the decoherence that scrambles other states, and hence they are exactly the states that support Quantum Darwinism \cite{Zurek2009NaturePhys,ZwolakZurek2013SRep,Duruisseau2023Entropy,Touil2024Quantum,Doucet2024PRX}. 

A simple estimate for how quickly a quantum system relaxes towards these pointer states is given by the decoherence rate. However, a more practically relevant question might be how quickly an observer will perceive a quantum system to be classical \cite{Kiely2026PRA}. More precisely, one is then interested in bounding the rate with which  the expectation value of a suitable observable approaches its pointer value. Importantly, this question is not only restricted to situations that will eventually support classical objectivity. Rather, quite generally it would be useful to know how quickly the outcome of a measurement can be determined ``with certainty''. 

A natural tool kit to address such a question is provided by \emph{quantum speed limits}, which impose fundamental restrictions on the rate of quantum evolution. Beginning with the Mandelstam-Tamm and Margolus-Levitin bounds, quantum speed limits have been developed through energetic, geometric, and information-theoretic approaches~\cite{MandelstamTamm1945, MargolusLevitin1998, TaddeiEtAl2013PRL, PiresEtAl2016PRX, DeffnerCampbell2017JPA, Deffner2017NJP, CampaioliEtAl2018PRL, Tripathy2026PRA}. 
While conventional formulations quantify evolution through a distance or distinguishability between quantum states~\cite{TaddeiEtAl2013PRL, PiresEtAl2016PRX, DeffnerCampbell2017JPA, Deffner2017NJP, CampaioliEtAl2018PRL, AnandanAharonov1990PRL, GiovannettiLloydMaccone2003PRA, DelCampoEtAl2013PRL, Deffner2013}, many physical tasks are specified not by the transformation of the complete state, but by a prescribed change in a particular measurable quantity~\cite{GarciaPintos2022PRX, MohanPati2022PRA, Hamazaki2022PRXQ,BudiyonoMoody2026PRA,budiyono2026quantumspeedlimitmeasurement}.

This has motivated expressing quantum speed limits directly in terms of observables and their expectation values~\cite{MohanPati2022PRA, GarciaPintos2022PRX, Hamazaki2022PRXQ, Shrimali2024PRA}. 
Such formulations distinguish dynamical changes that are operationally relevant to a chosen observable from changes in the quantum state that may leave that observable unaffected. 
Observable speed limits have subsequently been investigated through geometric constructions, many-body extensions, and their respective experimental implementations~\cite{BringewattEtAl2025PRR, MiaoEtAl2025SciAdv}. Related bounds on operator flows and correlation functions further demonstrate that observable speed provides a complementary perspective to state-based quantum speed limits~\cite{CarabbaHornedalDelCampo2022Quantum}.

The extension of quantum speed limits to open quantum systems has shown that environmental interactions can substantially modify both the accessible dynamical trajectories and the corresponding speed limits~\cite{TaddeiEtAl2013PRL, DelCampoEtAl2013PRL, DeffnerLutz2013PRL, CampaioliPollockModi2019Quantum, FunoShiraishiSaito2019NJP}.
While environmental memory can shorten state-based quantum speed-limit times or produce environment-assisted acceleration in particular models~\cite{DeffnerLutz2013PRL, CimmarustiEtAl2015PRL}, this does not imply that non-Markovianity universally accelerates every dynamical quantity. For our present purposes it is particularly remarkable that  there is evidence that non-Markovianity can hinder the  onset of classical objective reality~\cite{Giorgi2015PRA, Galve2016SRep, Milazzo2019PRA}, but that this is not always true~\cite{Oliveira2019PRA, Jahromi2020PRA}. 

In the present work, we reconcile these apparent contradictions by comprehensively analyzing the quantum speed limit for observables in the presence of Markovian and non-Markovian environments. As a main result, we find instances in which non-Markovianity both accelerates and delays the einselection of an observable. However,  when einselection becomes permanent, any delay is bounded from below by the accumulated negativity of the canonical decoherence rate. 

\section{Observable motion and quantum speed limits}
\label{sec:observable-speed}

\subsection{Observable motion under open quantum dynamics}
We consider a finite-dimensional system whose reduced dynamics is described by a differentiable family of completely positive and trace-preserving (CPTP) maps.
On intervals for which the dynamical maps are invertible, the evolution admits a regular time-local generator,
\begin{equation}
    \dot{\varrho}(t) = \mathcal{L}_t[\varrho(t)] = - i\,[H(t),\varrho(t)] + \mathcal{D}_t[\varrho(t)], 
    \label{eq:time-local-master}
\end{equation}
where the dot denotes the derivative with respect to time. 
The dissipator can be written in canonical form,
\begin{equation}
\begin{split}
    \mathcal{D}_t[\varrho]=\sum_\mu \gamma_\mu(t)\left(L_\mu(t)\varrho L_\mu^\dagger(t) -
        \frac{1}{2} \left\{ L_\mu^\dagger(t)L_\mu(t),\varrho\right\}
    \right),
    \label{eq:canonical-dissipator}
\end{split}
\end{equation}
where the Lindblad operators $L_\mu(t)$ form a traceless orthonormal operator set with respect to the Hilbert-Schmidt inner product, $\tr{L_\mu^\dagger(t)L_\nu(t)} = \delta_{\mu\nu}$, and $\gamma_\mu(t)$ are the canonical decoherence rates~\cite{GoriniKossakowskiSudarshan1976JMP, Lindblad1976CMP,
BreuerPetruccione2002, RivasHuelga2012}. 

In the simplest case, $\gamma_\mu(t)$ become time-independent and positive, in which case the dynamics represents a time-homogeneous semigroup case. For our present purposes, we are interested in the more general case in which the canonical decoherence operators $L_\mu(t)$ and rates $\gamma_\mu(t)$ may depend explicitly on time, and the rates need not remain nonnegative.  
The dynamics is said to be CP-divisible (completely positive divisible) on a given time interval if the time evolution map between any two instants within the interval exists and is completely positive (and trace-preserving) so that the dynamics factorizes into CP-intermediate maps~\cite{RivasHuelgaPlenio2010PRL, RivasHuelgaPlenio2014RPP}. 
In this time-local setting, the dynamics is CP-divisible if and only if $ \gamma_\mu(t)\geq 0 $ for every channel $\mu$ and every time $t$ in that interval~\cite{HallCresserLiAndersson2014PRA}. 
A strictly negative canonical decoherence rate $\gamma_\mu(t)<0$ therefore signals a breakdown of CP-divisibility.

In the following, we will use CP-indivisibility as a synonym for non-Markovianity~\cite{RivasHuelgaPlenio2010PRL, RivasHuelgaPlenio2014RPP, HallCresserLiAndersson2014PRA}.
This is distinct from explicit information-backflow formulations that instead diagnose non-Markovianity effects through revivals in the distinguishability of suitable pairs of states~\cite{BreuerLainePiilo2009PRL, LainePiiloBreuer2010PRA, BreuerLainePiiloVacchini2016RMP, Campbell2019NJPhys, Mendonca2024PRA}, although both notions capture the idea that non-Markovianity is associated with some flow of information from the environment to the system~\cite{Li2018PhysRep}, even if they do not always coincide~\cite{Chruscinski2011PRA}. 
For instance, this backflow of information can be harnessed as a quantifiable \textit{resource}, namely for generating entanglement in an open quantum system coupled to a finite, low-dimensional environment~\cite{Mirkin2019PRAEntanglingDynamics, Mirkin2019PRAInformationEntanglement}.

Our starting point is the definition of the instantaneous speed of an observable,
\begin{equation} \label{eq:speed-definition}
v_K(t) = \frac{1}{2} \left| \frac{d}{dt}\la K\ra_t \right|,
\end{equation}
where $\la K\ra_t = \tr{\varrho(t)K}$ is the expectation value of observable $K$ for the time-dependent state $\varrho(t)$ in a finite-dimensional Hilbert space $\mathcal{H}$.
For unitary \textit{closed}-system dynamics, this speed is upper-bounded as \cite{BudiyonoMoody2026PRA}
\begin{equation} \label{eq:trace-norm-asymmetry}
    A_K[\varrho(t)] = \frac{1}{2} \left\|[\varrho(t),K] \right\|_1, 
\end{equation}
where $\norm{\cdot}_1$ denotes the trace (Schatten-1) norm. $A_K[\varrho(t)]$ is the trace-norm asymmetry of the state $\varrho(t)$ relative to the group of translation unitaries generated by the observable $K$~\cite{MarvianSpekkens2016PRA, MarvianSpekkensZanardi2016PRA, MondalDattaSazim2016PLA, CampaioliYuPollockModi2022NJP} which can be operationally interpreted as the extremum imaginarity of the weak value of $K$ with the preselected state $\varrho(t)$ \cite{Budiyono2023PRA,Budiyono_2023asymmetrystrangeweakvalue}. For closed unitary dynamics, the trace-norm asymmetry $A_K[\varrho(t)]$ has been shown to be a genuine quantum resource for the quantum speed of observable~\cite{BudiyonoMoody2026PRA}. Moreover, $A_K[\varrho(t)]$ is a measure of the coherence in the state relative to the eigenbasis of $K$. Hence, for open-system dynamics, the observable speed can be understood as arising from the conversion of this available coherence through coherent dynamics~\cite{BudiyonoMoody2026PRA}, together with dissipative contributions from the environment~\cite{GarciaPintos2022PRX}. 
 
For \textit{open}-system dynamics~\eqref{eq:time-local-master}, the observable motion separates exactly into \textit{coherent} and \textit{dissipative} contributions. Using the adjoint relation (see Appendix~\ref{app:adjoint-relation}), we obtain
\begin{equation}
\frac{d}{dt} \la K\ra_t = - i\,\tr{H(t)[\varrho(t),K]} + \la \mathcal{D}_t^\dagger[K] \ra_t,
\label{eq:observable-speed}
\end{equation}
where $\mathcal{D}_t^\dagger$ is the adjoint dissipator, whose action on an observable $K$ is
\begin{equation}
    \mathcal{D}_t^\dagger[K] = \sum_\mu \gamma_\mu(t) \left( L_\mu^\dagger(t)K L_\mu(t) -  \frac{1}{2} \left\{ L_\mu^\dagger(t)L_\mu(t),K \right\} \right). 
    \label{eq:adjoint-dissipator}
\end{equation} 
We refer to the two terms contributing to the rate of change of $\la K\ra_t$, as the coherent and dissipative observable currents, 
\begin{align}
    J_K^\mathrm{coherent}(t) &\equiv - i\,\tr{H(t)[\varrho(t),K]} \label{eq:coherent-current} \\
    J_K^\mathrm{dissipative}(t) &\equiv \la \mathcal{D}_t^\dagger[K] \ra_t, \label{eq:dissipative-current}
\end{align}
so that the instantaneous observable speed can be expressed in terms of the magnitude of the total current, 
\begin{equation}
v_K(t) = \frac{1}{2} \left| J_K^\mathrm{coherent}(t) + J_K^\mathrm{dissipative}(t) \right|. 
\label{eq:instantaneous-speed-currents}
\end{equation}
The coherent current requires noncommutativity between the state and the observable, and hence vanishes whenever $[\varrho(t),K]=0$. 
The dissipative current is instead determined by the action of the adjoint dissipator on the observable, independent of $[\varrho(t),K]$.  

This distinction does not exhaust the environmental influence. 
Dissipation also modifies the solution $\varrho(t)$ and can thereby suppress, preserve, or restore the asymmetry entering the coherent current. 
We therefore distinguish between a direct dissipative effect, through $J_K^\mathrm{dissipative}(t)$, and an indirect dissipative effect, through the dissipative modification of $\varrho(t)$, and hence of $[\varrho(t),K]$. 
This decomposition complements previous analyses of observable speed and connections between asymmetry, bath-induced fluctuations, and expectation-value currents~\cite{GarciaPintos2022PRX,SekiguchiEtAl2024arXiv}. 

The sign of the canonical decoherence rate $\gamma_\mu(t)$ determines the signed contribution of the corresponding channel to the dissipative current. 
A reversal of this sign reverses the dissipative current of the $\mu$th channel, but does not alone imply an overall enhancement of the observable speed~\eqref{eq:instantaneous-speed-currents}, since this may reinforce or oppose the coherent current and contributions from other dissipative channels.

\subsection{Instantaneous speed limit} 
\label{sec:instantaneous-bound}
We first establish the instantaneous bound underlying the subsequent analysis for some fixed observable $K=K^\dagger$. 
Without a constraint on the resources available to the quantum system, the coherent current can be made arbitrarily large by rescaling the Hamiltonian. 
A meaningful speed limit, therefore, requires an appropriate normalization. 
We fix the scale of the coherent driving by imposing unit operator norm (equivalent to the Schatten-$\infty$ norm) $\norm{H(t)}_\infty=1$.  
Applying the triangle inequality to Eqs.~\eqref{eq:speed-definition} and \eqref{eq:observable-speed}, 
\begin{equation} \label{eq:triangle-inequality}
v_K(t) \leq \frac{1}{2} \left|-i\,\tr{H(t)[\varrho(t),K]} \right| + \frac{1}{2} \left| \la\mathcal{D}_t^\dagger[K]\ra_t \right|,
\end{equation}
where the coherent current satisfies the Schatten-H\"{o}lder inequality (Appendix~\ref{app:observable-currents-bounds})
\begin{equation} \label{eq:schatten-holder}
\left|-i\,\tr{H(t)[\varrho(t),K]} \right| \leq \norm{H(t)}_\infty \norm{[\varrho(t),K]}_1.
\end{equation}
This yields the instantaneous observable-speed bound
\begin{align}
v_K(t) & \leq \frac{1}{2} \left(\norm{[\varrho(t),K]}_1 + \left| \la\mathcal{D}_t^\dagger[K]\ra_t \right| \right)
\nonumber \\
&\leq A_K[\varrho(t)] + \frac{1}{2} \left| \la\mathcal{D}_t^\dagger[K]\ra_t \right|, 
\label{eq:instantaneous-bound}
\end{align}
where we have used the definition of the trace-norm asymmetry~\eqref{eq:trace-norm-asymmetry}. 
This is our first main result. 
The two terms are not independent, and enter the bound asymmetrically: the first term is a \textit{capacity} in that it quantifies the asymmetry available for coherent motion, equal to the largest coherent current attainable by any admissible $H(t)$, whereas the second is the realized current of the fixed physical dissipator, for which no analogous resource normalization applies.

We emphasize again that, in general, the two terms in Eq.~\eqref{eq:instantaneous-bound} are \textit{not} additive contributions to the actual speed~\eqref{eq:instantaneous-speed-currents}. Importantly, the coherent contribution saturates the Schatten-H\"{o}lder inequality~\eqref{eq:schatten-holder} when $H(t)$ is aligned with the sign operator of $-i\,[\varrho(t),K]$ (Appendix~\ref{app:observable-currents-bounds}). Saturation of the complete bound~\eqref{eq:instantaneous-bound} additionally requires constructive alignment of the coherent and dissipative currents~\eqref{eq:instantaneous-speed-currents}. 

\subsection{Amplitude-damping channel}
These conditions can be realized for amplitude-damping dynamics, namely
\begin{align} \label{eq:amplitude-damping}
\dot{\varrho}(t) &= -i\,[H,\varrho(t)]+
\gamma\left(\sigma_-\varrho(t)\sigma_+ - \frac{1}{2}\{ \sigma_+\sigma_-, \varrho(t) \} \right),
\end{align}
with constant decoherence rate $\gamma>0$, unit-norm Hamiltonian $\norm{H}_\infty=1$, and Hilbert-Schmidt normalized jump operator $L=\sigma_-= \ket{0}\!\bra{1}$. 
For the initially excited state $\varrho(0)=\ket{1}\!\bra{1}$ and a unit-norm Hamiltonian commuting with $\sigma_z$, $[H,\sigma_z]=0$, the state remains diagonal and the solution is
\begin{equation} \label{eq:amplitude-damping-solution}
\varrho(t) = \begin{pmatrix}
1-e^{-\gamma t} & 0 \\
0 & e^{-\gamma t}
\end{pmatrix}.
\end{equation}
If we choose observable $K=\sigma_z$, since $\varrho(t)$ and $K$ remain diagonal in the same basis, $[\varrho(t),K]=0$ and the coherent current vanishes for all $t\geq0$. 
The observable motion is therefore generated entirely by the dissipative current (Appendix~\ref{app:amplitude-damping}), 
\begin{equation} \label{eq:amplitude-damping-speed}
v_K(t) = \gamma e^{-\gamma t} 
= \frac{1}{2} \left| \la \mathcal{D}_t^\dagger[K] \ra_t \right|,
\end{equation}
so the instantaneous observable-speed bound is saturated at all times. 

\section{Dissipative currents and pointer observables}
\label{sec:dissipative-currents-pointer-observables} 

The instantaneous observable speed~\eqref{eq:instantaneous-speed-currents} is determined by a coherent current~\eqref{eq:coherent-current} generated by the Hamiltonian and a dissipative current~\eqref{eq:dissipative-current} generated directly by the environment. 
To identify the observables for which the latter is relevant, we examine the structure of the adjoint dissipator~\eqref{eq:adjoint-dissipator}.  

The channel structure becomes particularly transparent in the commutator representation
\begin{equation}
    \mathcal{D}_t^\dagger[K] = \frac{1}{2} \sum_\mu \gamma_\mu(t) \left( L_\mu^\dagger(t)[K,L_\mu(t)]+ [L_\mu^\dagger(t),K]L_\mu(t) \right),
\label{eq:commutator-dissipator}
\end{equation}
derived in Appendix~\ref{app:adjoint-commutator-form}. 
This makes explicit that the direct dissipative action on $K$ depends on its commutation relations with the decoherence operators. 
In particular, a sufficient condition for the $\mu$th channel to leave $K$ unaffected directly is
\begin{equation}
    [K,L_\mu(t)] = [K,L_\mu^\dagger(t)] = 0.
    \label{eq:pointer-commutation}
\end{equation}
If this condition holds for every channel, then $\mathcal{D}_t^\dagger[K]=0$, and the environment produces no direct dissipative current in $K$. 
Interestingly, the action of the adjoint dissipator can also be interpreted in terms of the asymmetry of the decoherence operators with respect to the observable $K$ as a resource for quantum speed, cf. Appendix~\ref{app:adjoint-commutator-form-asymmetry}.
 
Equation~\eqref{eq:pointer-commutation} therefore identifies observables on which the dissipator has no direct action, $\mathcal{D}^\dag[K] = 0$.
Such observables are compatible with the environmental monitoring and provide natural candidates for \textit{pointer observables} \cite{Zurek2025Book}, whose expectation values are immune to environmental decoherence. 
Any residual motion of a pointer observable is then purely Hamiltonian, not environmental. 

Conversely, when $\mathcal{D}_t^\dagger[K]\neq0$, the environment can directly drive the expectation value of $K$. 
In this case, the observables are instead natural candidates for \textit{coherence witnesses}, whose expectation values can be driven directly by the environment toward their pointer values $K_\infty = \lim_{t\to\infty} \la K\ra_t$, i.e., $\la K \ra$ evaluated on the einselected (diagonal) state the system relaxes to.  
However, this direct action does not exhaust the influence of the environment: even when $\mathcal{D}_t^\dagger[K]=0$, the dissipator can modify the state $\varrho(t)$ and thereby change the asymmetry~\eqref{eq:trace-norm-asymmetry} available for coherent motion. 

\subsection{Pure-dephasing channel} 
To illustrate the interplay between coherent and dissipative
contributions, consider a time-homogeneous pure-dephasing semigroup, 
\begin{equation} \label{eq:pure-dephasing}
    \dot{\varrho}(t) = -i\,[\sigma_z,\varrho(t)] +
    \frac{\gamma}{2} \left(\sigma_z\varrho(t)\sigma_z -\varrho(t)\right), 
\end{equation}
with constant decoherence rate $\gamma>0$ and $L={\sigma_z}/{\sqrt{2}}$, so that $\tr{L^\dagger L}=1$. 
We take the initial state $\varrho(0)=\ket{+}\!\bra{+}$ and choose the observable (coherence witness) $K=\sigma_x$. 
For these choices in this time-homogeneous case, 
\begin{equation} \label{eq:adjoint-dissipator-sigmax}
    \mathcal{D}^\dagger[\sigma_x] = -\gamma\sigma_x,
\end{equation} 
so that the environment directly drives the coherence witness $K=\sigma_x$. 
Together with the noncommuting Hamiltonian in Eq.~\eqref{eq:pure-dephasing}, this produces competing coherent and dissipative currents.
The explicit solution of the master equation and the evaluation of the terms entering the instantaneous observable-speed bound~\eqref{eq:instantaneous-bound} are given in Appendix~\ref{app:pure-dephasing}.

\begin{figure}
\centering
\includegraphics[width=\columnwidth]{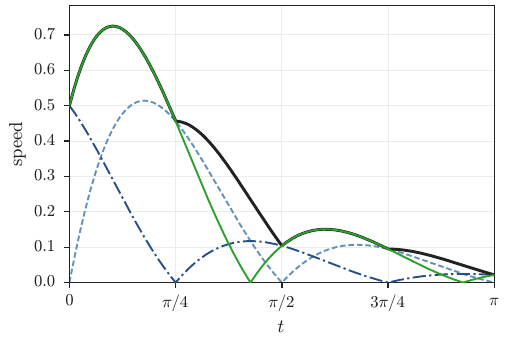}
\caption{
Instantaneous observable speed for the pure-dephasing dynamics with $H=\sigma_z$, $L=\sigma_z/\sqrt{2}$, $\gamma=1$, $\varrho(0)=\ket{+}\!\bra{+}$, and $K=\sigma_x$.
The thin solid green curve shows the exact speed $v_K(t)$, while the thick solid black curve shows the instantaneous observable-speed bound.
The dash-dotted dark-blue and dashed mid-blue curves show the dissipative contribution $\frac{1}{2}\left|\la\mathcal{D}^{\dagger}[K]\ra_t\right|$ and the coherent contribution $\frac{1}{2}\norm{[\varrho(t),K]}_1$, respectively.
}
\label{fig:pure-dephasing-speed}
\end{figure}

Figure~\ref{fig:pure-dephasing-speed} illustrates the tightness of the instantaneous speed bound~\eqref{eq:instantaneous-bound}. 
For this solution, the coherent contribution saturates the Schatten-H\"{o}lder inequality at every instant, so any separation between $v_K(t)$ and the instantaneous bound arises from the triangle inequality. 
When the coherent and dissipative currents have the same sign, they reinforce one another and the bound is saturated; when their signs differ, their partial or complete cancellation yields a strict inequality. 
In particular, $v_K(t)$ can vanish while both currents remain nonzero. 
Thus, reinforcement and cancellation of observable currents occur even for a constant positive dephasing rate, i.e., within a CP-divisible quantum dynamical semigroup.

This example shows that the observable speed is controlled not only by the magnitude of the coherent and dissipative currents, but also by their relative sign.
Non-Markovianity can modify the speed of a coherence witness either directly, through the dissipative contribution to Eq.~\eqref{eq:observable-speed}, or indirectly, by modifying the state and thereby restoring or suppressing the asymmetry available to the coherent contribution. 
The following section examines these two mechanisms in the presence of non-Markovian rate reversals and their consequences for the speed and permanence of einselection.

\subsection{Einselection at the observable speed limit}
\label{sec:einselection-speed-limit}
We now ask when the instantaneous observable-speed bound~\eqref{eq:instantaneous-bound} can be attained during einselection, and identify the conditions under which a coherence witness can reach its pointer value at the maximum allowed rate.
The speed limit is particularly informative when $K$ is chosen as a coherence witness, since the dissipative dynamics can then drive $\la K\ra_t$ toward its pointer value.

We first consider a configuration in which the coherent contribution vanishes and the bound is attained entirely through the dissipative current by isolating the dissipative contribution using the \textit{pure-dephasing} model~\eqref{eq:pure-dephasing} introduced in the previous section. 
With $L={\sigma_z}/{\sqrt{2}}$, $K=\sigma_x$, $\varrho(0)=\ket{+}\!\bra{+}$, we now choose the Hamiltonian to be the unit-norm inert generator $H=\openone$. 
This choice eliminates the coherent current while satisfying the normalization $\|H\|_\infty=1$ used to define the speed limit.
Since the initial state is diagonal in the eigenbasis of $K$ and the pure-dephasing dynamics preserves this basis, $[\varrho(t),K]=0$ throughout the evolution. 
The asymmetry available for coherent motion therefore vanishes, $A_K[\varrho(t)]=0$, and the observable motion is generated entirely by the dissipative current.

For this channel, Eq.~\eqref{eq:adjoint-dissipator-sigmax} holds, so that $\la K\ra_t=\la K\ra_0 e^{-\gamma t}=e^{-\gamma t}$. 
At the same time,
\begin{equation}
    \frac{1}{2} \left| \la \mathcal{D}^\dagger[K] \ra_t \right|
    = \frac{\gamma}{2}e^{-\gamma t} = v_K(t).
\end{equation}
Thus, the instantaneous observable-speed bound~\eqref{eq:instantaneous-bound} is saturated at every time. 
Hence, for this configuration, \textit{einselection takes place at the observable speed limit}: the dissipative current alone realizes the maximum observable speed allowed by Eq.~\eqref{eq:instantaneous-bound}.
Here, the witness \textit{relaxes} toward its pointer value, with its speed decaying as it approaches it, which is the case whose delay under memory we bound in Sec.~\ref{sec:non-markovian-delay}.

This saturation is not specific to pure dephasing. 
The amplitude-damping model~\eqref{eq:amplitude-damping} and corresponding solution~\eqref{eq:amplitude-damping-solution} provide a second realization in which the observable is diagonal in the pointer basis and its motion is again generated entirely by the dissipative current, which alone drives the witness towards its pointer value. 
The two examples establish that speed-limited einselection is a structural property of the observable-speed bound rather than a special feature of a particular dissipative channel. 

More generally, however, saturation does not require the coherent contribution to vanish. 
At the instantaneous level, the observable speed~\eqref{eq:instantaneous-speed-currents} is equal to the observable-speed limit~\eqref{eq:instantaneous-bound} only when the coherent current saturates its Schatten–H\"{o}lder bound and the two currents have compatible signs. 

These conditions provide the reference point for the non-Markovian analysis that follows. 
Whether memory enhances or suppresses einselection is then determined by how the direct and indirect mechanisms identified above drive the currents toward or away from the saturation conditions.

\section{Non-Markovian modification of einselection}
\label{sec:non-markovian-modification}

We next examine how non-Markovianity modifies the observable currents, and therefore the observable motion.
We take non-Markovianity in the sense of CP-indivisibility, which in a time-local description is signalled by negative canonical decoherence rates. 

\subsection{CP-indivisibility and reversal of dissipative currents}
For the canonical time-local generator, CP-divisibility requires $\gamma_\mu(t)\geq0$ for all channels and times. 
A negative canonical decoherence rate therefore signals a breakdown of CP-divisibility at that time~\cite{RivasHuelgaPlenio2010PRL, HallCresserLiAndersson2014PRA, RivasHuelgaPlenio2014RPP}. 
In the observable current, however, the relevant quantity is not the sign of $\gamma_\mu(t)$ alone but its signed contribution,
\begin{align} \label{eq:channel-resolved-dissipative-current}
    &\left( J^\mathrm{dissipative}_{K}(t) \right)_{\mu} \nonumber \\
    &\quad = \gamma_\mu(t) \la L_\mu^\dagger(t) K L_\mu(t) - \frac{1}{2} \{L_\mu^\dagger(t) L_\mu(t),K \} \ra_t. 
\end{align}

Thus, for fixed $L_\mu(t)$ and $\varrho(t)$, changing the sign of $\gamma_\mu(t)$ reverses the signed current generated by that channel. 
The observable speed, however, depends on the absolute value of the total current and therefore on its alignment with the coherent current and with any other dissipative contributions.  

To examine the effect of rate reversal, we consider a \textit{single} dissipative channel and compare the non-Markovian (NM) evolution with a matched CP-divisible (Markovian, M) reference having the same instantaneous rate magnitude,
\begin{equation}
    \gamma_{\mathrm{M}}(t)
    =
    \left|\gamma_{\mathrm{NM}}(t)\right|.
    \label{eq:matched-cp-rate}
\end{equation}
The two evolutions therefore differ only in the sign of the canonical decoherence rate during the CP-indivisible intervals. 
Define the accumulated canonical-rate negativity by
\begin{equation} \label{eq:accumulated-negativity}
    F(0,t) \equiv \int_0^t ds\,[-\gamma_{\mathrm{NM}}(s)]_+, \qquad [x]_+\equiv\max\{x,0\}, 
\end{equation}
which equals half the difference between the integrated canonical decoherence rates $\Gamma_i(0,t)\equiv\int_0^t ds\,\gamma_i(s)$ (for $i=\mathrm{M},\mathrm{NM}$) of the matched pair over the interval $[0,t]$,
\begin{equation}
\begin{split}
    2F(0,t)&= \Gamma_{\mathrm{M}}(0,t) - \Gamma_{\mathrm{NM}}(0,t) \\
    &= \int_0^t ds \left(|\gamma_{\mathrm{NM}}(s)| - \gamma_{\mathrm{NM}}(s) \right)
\end{split}
    \label{eq:Gamma-difference}
\end{equation}
For a single canonical channel, $F(0,t)$ is the integrated negative part of the canonical decoherence rate. 
For a Hilbert-Schmidt normalized jump operator, $\tr{L^\dagger L}=1$, this coincides with the (finite-time) Rivas-Huelga-Plenio measure of CP-indivisibility~\cite{RivasHuelgaPlenio2010PRL} accumulated on $[0,t]$ using the canonical-rate form of the measure~\cite{HallCresserLiAndersson2014PRA, RivasHuelgaPlenio2014RPP}. We therefore adopt this normalization, and otherwise note that a different choice simply scales the accumulated negativity by a positive constant, which then leaves its sign unchanged.
This single-channel construction isolates the effect of rate reversal and will be used below to distinguish direct and indirect modifications of observable motion.

\subsection{Direct enhancement of dissipative observable motion}
\label{sec:non-markovian-modification-direct-enhancement}
The first mechanism arises in a single-channel amplitude-damping model~\eqref{eq:amplitude-damping} with $K=\sigma_z$ and an initially excited state. 
Since $[\varrho(t),K]=0$, the coherent current vanishes identically and the observable motion is generated entirely by the dissipative current.
The population entering the dissipative current is proportional to the corresponding exponential decay factor. 
The relation~\eqref{eq:Gamma-difference} gives
\begin{align}
    \varrho_{11}^{\mathrm{NM}}(t)
    &=
    e^{2F(0,t)}
    \varrho_{11}^{\mathrm{M}}(t), \label{eq:nm-direct-enhancement-rho11} \\
    v_K^{\mathrm{NM}}(t)
    &=
    e^{2F(0,t)}
    v_K^{\mathrm{M}}(t).
    \label{eq:nm-direct-enhancement-speed}
\end{align}
Since $F(0,t)\geq0$, the non-Markovian evolution has a speed at least as large as that of the matched CP-divisible reference, with a strict enhancement whenever $F(0,t)>0$ and the instantaneous current is nonzero.
The enhancement does not arise from a larger rate magnitude, since $|\gamma_{\mathrm{NM}}(t)|=|\gamma_{\mathrm{M}}(t)|$. Instead, negative-rate intervals retain or restore excited-state population, leaving more population available for the dissipative current. 

On negative intervals ($\gamma_{\mathrm{NM}}(t)<0$), this enhanced speed corresponds to population flowing back into the excited state, so the witness momentarily moves \textit{away} from its pointer value; the net approach is treated in Sec.~\ref{sec:permanent-transient-einselection}. 
In this model, CP-indivisibility therefore \textit{enhances} the instantaneous observable speed relative to the matched CP-divisible reference directly through the dissipative channel. 
The result is not universal: for a general observable, the reversed channel current may instead oppose the coherent or other dissipative currents and \textit{reduce} the total instantaneous observable speed.

\subsection{Indirect enhancement of observable motion through asymmetry restoration}
\label{sec:non-markovian-modification-indirect-enhancement}
The complementary situation is obtained when the dissipator leaves the pointer observable invariant, so that the dissipative current vanishes identically. 
Non-Markovianity can nevertheless affect the observable speed by modifying the trace-norm asymmetry available to the coherent current.

Consider pure dephasing with the pointer observable $K_{\mathrm{ptr}}=\sigma_z$, and a decoherence operator $L = \sigma_z/\sqrt{2}$. 
Since $[K_{\mathrm{ptr}},L]=0$, the dissipative current of the pointer observable vanishes. 
Instead of directly moving the pointer observable, the environment modifies the state and hence the asymmetry $A_{\sigma_z}[\varrho(t)]=\frac{1}{2} \left\| [\varrho(t),\sigma_z] \right\|_1$ available for coherent motion. 
For a qubit and the pointer observable $K_{\mathrm{ptr}}=\sigma_z$, this is equal to the $\ell_1$ coherence in the pointer basis $\{\ket{k}\}$ \cite{Budiyono2023PRA},
\begin{equation}
    \frac{1}{2} \left\| [\varrho(t),\sigma_z] \right\|_1 
    = C_{\ell_1}^{(\mathrm{ptr})}[\varrho(t)]
   = \sum_{k \neq k'} \left| \bra{k} \varrho(t) \ket{k'}\right|. 
    \label{eq:pointer-asymmetry-coherence}
\end{equation}
Thus, in this realization, coherence provides a direct representation of the state resource that limits the coherent observable current. Since the pointer-basis coherence is proportional to the corresponding dephasing factor, for a matched pair~\eqref{eq:matched-cp-rate} of pure-dephasing dynamics, one obtains 
\begin{equation} \label{eq:nm-coherence-enhancement}
    C_{\ell_1}^{\mathrm{NM}}(t) = e^{2F(0,t)} C_{\ell_1}^{\mathrm{M}}(t).
\end{equation}
The same enhancement therefore appears in the maximal coherent observable current. 
For a unit-norm Hamiltonian saturating the Schatten-H\"{o}lder bound,
\begin{equation}  \label{eq:nm-indirect-enhancement}
    v_{K_{\mathrm{ptr}}}^{\mathrm{NM}}(t) = e^{2F(0,t)} v_{K_{\mathrm{ptr}}}^{\mathrm{M}}(t).
\end{equation}

\begin{figure}
    \centering
    \includegraphics[trim=19cm 1cm 19cm 2cm, clip=true, width=\columnwidth]{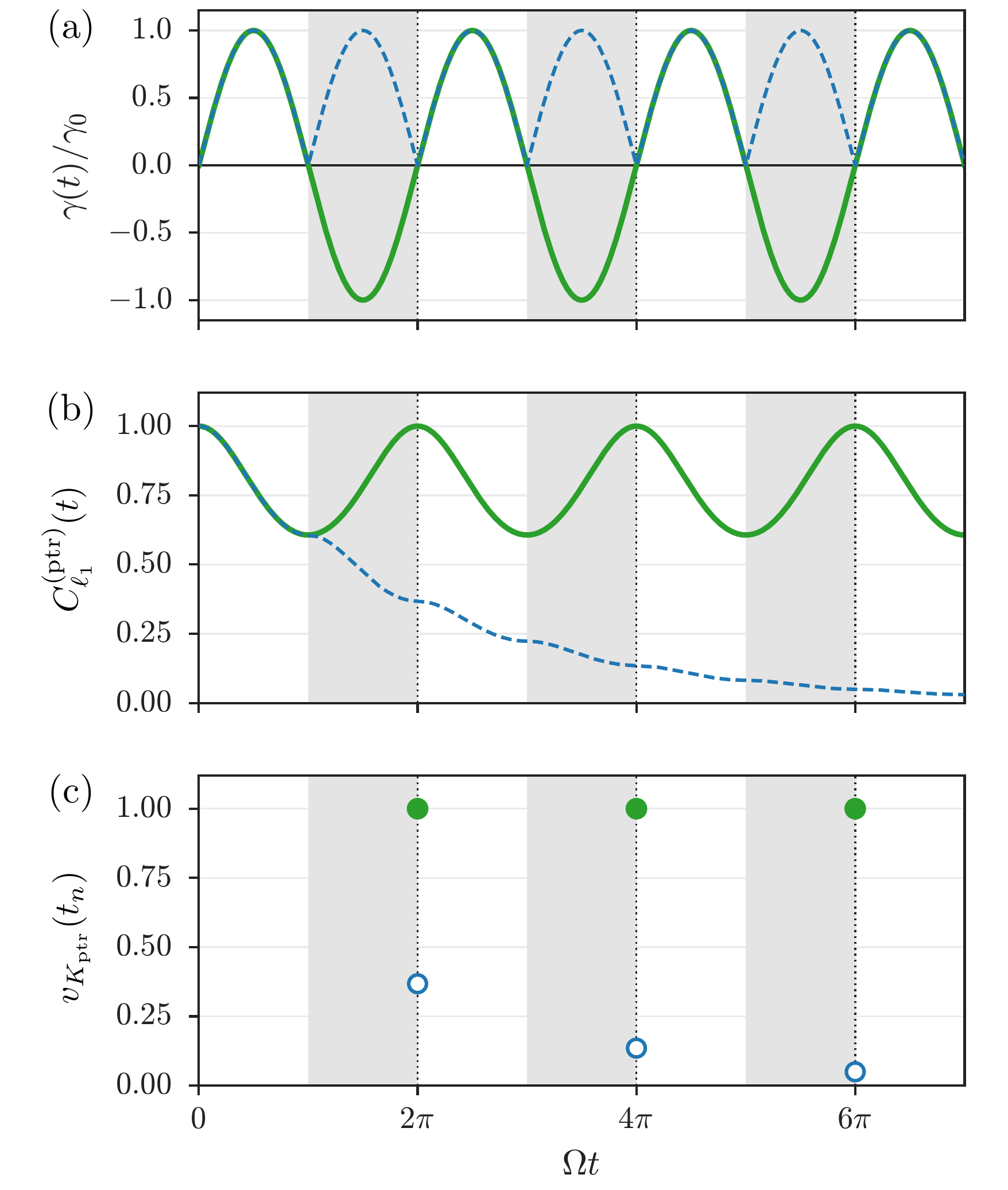}
    \caption{{Matched CP-indivisible and CP-divisible pure-dephasing dynamics for the pointer observable $K_{\mathrm{ptr}}=\sigma_z$. 
    Solid green curves denote the CP-indivisible evolution generated by $\gamma_{\mathrm{NM}}(t)=\gamma_0\sin{(\Omega t)}$ with $\gamma_0/\Omega=1/4$, while dashed blue curves denote the matched CP-divisible reference generated by $\gamma_{\mathrm{M}}(t)=|\gamma_{\mathrm{NM}}(t)|$. 
    Shaded regions indicate intervals for which $\gamma_{\mathrm{NM}}(t)<0$, and dotted vertical lines mark the comparison times $t_n=2\pi n/\Omega$ (see Appendix~\ref{app:nonmarkovian-bounds}). 
    (a) Canonical decoherence rates normalized by $\gamma_0$. 
    The rates coincide outside the shaded intervals. 
    (b) Pointer-basis $\ell_1$ coherence. 
    Under the CP-indivisible dynamics, the coherence undergoes complete revivals at the comparison times $t_n$, while it decays monotonically under the CP-divisible reference. 
    (c) Instantaneous pointer-observable speed evaluated at $t_n$, with the preparation Hamiltonian $H_p$ replaced by the normalized control drive $H_c(t_n)$. 
    During the preparation stage, $[H_p,K_{\mathrm{ptr}}]=0$ and $\mathcal{D}_t^\dagger[K_{\mathrm{ptr}}]=0$, so the pointer observable remains stationary. 
    With the chosen normalization, the switched drive saturates the coherent contribution to the observable-speed bound, giving $v_{K_{\mathrm{ptr}}}^{\mathrm{NM}}(t_n)=1$, while $v_{K_{\mathrm{ptr}}}^{\mathrm{M}}(t_n)=e^{-4n\gamma_0/\Omega}$.
    }
    }
    \label{fig:pointer-coherence}
\end{figure}

Figure~\ref{fig:pointer-coherence} illustrates this indirect mechanism.
Negative-rate intervals restore the pointer-basis coherence and hence the available trace-norm asymmetry. 
The normalized Hamiltonian drive converts this restored asymmetry into a larger coherent observable current. 
The detailed construction of the saturating Hamiltonian and the corresponding qubit calculation are given in Appendix~\ref{app:nonmarkovian-bounds}.
In contrast to the amplitude-damping example, the enhanced speed in Eq.~\eqref{eq:nm-indirect-enhancement} is not carried by a dissipative current of the pointer observable, which vanishes identically.
Non-Markovianity instead restores the trace-norm asymmetry and thereby increases the coherent current that can be generated from the state. 
The enhancement is consequently indirect: memory first modifies the state and only then changes the observable speed through the coherent current.

It is important to distinguish the negative-rate description from the physicality of the reduced dynamics. 
A time-local generator can be CP-indivisible while the resulting dynamical map remains completely positive; a negative canonical decoherence rate therefore does not imply that the reduced state solution becomes unphysical. 
In the single-channel models considered here, $F(0,t)$ measures the accumulated canonical-rate negativity and vanishes for CP-divisible evolution.

The two mechanisms act on different components of the observable motion. 
In the amplitude-damping case, non-Markovianity enhances the population supporting the dissipative current. 
In the dephasing case, the dissipative current of the pointer observable vanishes, while memory restores the trace-norm asymmetry that limits the coherent current. 
Thus, the same accumulated rate negativity $F(0,t)$ can enhance observable motion through two distinct pathways: directly, by increasing the state-dependent dissipative current, or indirectly, by restoring the asymmetry available to the coherent current. 

\section{Permanent versus transient einselection}
\label{sec:permanent-transient-einselection}

The speed of a coherence witness does not determine whether einselection is permanent. 
For instance, a witness can move rapidly toward a pointer value, before moving away from it, and later reviving. 
Conversely, coherence can decay permanently even when the dynamics contains intervals of CP-indivisibility.
Einselection permanence is instead determined by the long-time behavior of the environmental decoherence function~\cite{JessRiedel2012NJP}. 

For the exactly solvable independent boson model~\cite{BreuerPetruccione2002, Mahan2000}, corresponding to the $\Delta=0$ limit of the spin-boson model~\cite{Leggett1987}, the coherence can be written as
\begin{equation} \label{eq:dephasing-coherence}
    \varrho_{01}(t) = \varrho_{01}(0) e^{-i\Phi(t)} e^{-\Gamma(t)}, 
\end{equation}
where $\Phi(t)$ is a bath-induced phase and $\Gamma(t)\geq 0$ is the decoherence function. 
For a bosonic environment initially in thermal equilibrium, 
\begin{equation} \label{eq:bath-spectral-density}
  \Gamma(t) = \int_0^\infty \, d\omega\,
    \frac{J(\omega)}{\omega^2}\,\coth\left(\frac{\omega}{2k_BT}\right)\,
    \left(1-\cos{\omega t}\right) 
\end{equation}
where $J(\omega) \sim \sum_k |g_k|^2 \delta(\omega-\omega_k)$ is the bath spectral density, whose normalization is fixed by the system-bath coupling convention, and where $g_k$ characterizes the coupling with the $k$th mode. 
If we assume $J(\omega) \geq 0$, the integrand is nonnegative for every $\omega$ and all times $t\geq0$, so $\Gamma(t)\geq 0$, too. 
For this model, the decoherence function coincides with the integrated canonical decoherence rate of Sec.~\ref{sec:non-markovian-modification}, $\Gamma(t)=\int_0^t ds\,\gamma(s)$, and the accumulated negativity of the canonical decoherence rate is the quantity $F(0,t)$ of Eq.~\eqref{eq:accumulated-negativity}.

Whether coherence is permanently suppressed is determined by the long-time behavior of $\Gamma(t)$, rather than by its monotonicity.
If 
\begin{equation}
    \lim_{t\to\infty}\Gamma(t) = \infty,\quad \mbox{then} \quad \lim_{t\to\infty}\left|\varrho_{01}(t) \right|= 0,
\end{equation}
and the corresponding coherence witness reaches its pointer value asymptotically, even when $\Gamma(t)$ approaches this limit nonmonotonically through intervals of transient recoherence, that is, through nonzero $F(0,t)$. 

At zero temperature, an Ohmic spectral density with exponential cutoff, $J(\omega) = A \omega e^{-\omega/\omega_c}$, gives $\Gamma(t) = \frac{A}{2}\ln{(1+\omega_c^2t^2)}$. 
Hence $\Gamma(t)\rightarrow \infty$ logarithmically, and the coherence decays algebraically,
\begin{equation}
    \left|\varrho_{01}(t)\right| \sim(\omega_ct)^{-A}. 
\end{equation}
so einselection is therefore permanent. 

If instead $\Gamma(t)$ remains bounded, a nonzero coherence amplitude survives asymptotically and complete einselection does not occur. 
For instance, Ref.~\cite{Strathearn2017NJP} solves the super-ohmic independent boson model exactly and finds that $\Gamma(t)$ converges to a finite plateau rather than diverging, where the approach to the plateau is nonmonotonic, with a transiently negative canonical rate $\gamma(t)$ (hence $F(0,t)>0$). 
This nonmonotonicity stems from the super-ohmic exponent of an otherwise featureless spectral density rather than any resonant structure. 

By contrast, a spectrally peaked environment can produce nonmonotonic $\Gamma(t)$, so that the accumulated decoherence is temporarily reduced and the coherence partially revives.
Within the pure-dephasing description~\eqref{eq:dephasing-coherence}, these revivals are encoded directly in the nonmonotonicity of $\Gamma(t)$, i.e., in a nonzero $F(0,t)$. 
Therefore, einselection can be transient even when the witness has previously approached its pointer value. 
A distinct mechanism manifests in dissipative structured environments, where coherence is lost through population decay and its revivals instead take place through resonant excitation exchange. 
In a Lorentzian environment, which encompasses the paradigmatic Jaynes-Cummings model~\cite{JaynesCummings1963} and its many variants~\cite{LarsonMavrogordatos2021} (see Appendix~\ref{app:microscopic-models} for some relevant choices), the corresponding coherence can undergo revivals~\cite{BreuerPetruccione2002, BreuerLainePiilo2009PRL, BreuerLainePiiloVacchini2016RMP}, which manifest in derived speed limits~\cite{Deffner2013, Zhang2014Srep}. 
The distinction between these behaviors is therefore determined by the long-time spectral structure, rather than by the sign of the instantaneous canonical decoherence rate alone. 

The two notions are governed by different properties of the same decoherence function, namely permanence by whether $\Gamma(t)$ diverges, and CP-divisibility by whether $\Gamma(t)$ is monotonic, i.e., by the sign of $\gamma(t)=\dot\Gamma(t)$. 
They are logically independent in the sense that a process can be CP-divisible and permanently decohering, CP-indivisible and permanently decohering, or CP-indivisible with transient einselection. 
In the pure-dephasing family, CP-divisibility forces $\Gamma(t)$ to be monotonic and hence precludes revivals, so CP-divisible processes with transient einselection do not arise, and transience necessarily requires CP-indivisibility. 

A particularly useful realization combines these two behaviors in a single environment (still at zero temperature). 
Consider the spectral density 
\begin{equation}\label{eq:structured-J}
  J(\omega) = {A\,\omega\,e^{-\omega/\omega_c}}
    + B\,\frac{\lambda \omega}{(\omega-\omega_0)^2+\lambda^2}, 
\end{equation}
where $A$ and $B$ are nonnegative and real, which describes, e.g., a qubit in an Ohmic background undergoing resonance with the quasimode $\omega_0$. The first term supplies the low-frequency Ohmic weight and therefore the long-time divergence $\Gamma(t)\to\infty$ required for permanent decoherence, while the second term introduces a Lorentzian resonance (halfwidth $\lambda$) that ensures nonmonotonicity of the rate $\gamma(t)$, thereby realizing the divergent but nonmonotonic $\Gamma(t)$ anticipated above. 
In the regime $\omega_c \gtrsim w_0 \gg \lambda$, the roles decouple cleanly, and smaller $\omega_0/\lambda$ gives increasingly long-lived (yet transient) revivals, i.e., larger accumulated negativity $F(0,t)$. 
Non-Markovianity characterizes the local divisibility of the reduced dynamics, whereas permanence characterizes its asymptotic approach to the pointer state.

\section{Non-Markovian delay of einselection}
\label{sec:non-markovian-delay}

Non-Markovianity can coexist with permanent einselection, but its effect on the rate of approach to the pointer value is not necessarily an acceleration. 
The CP-indivisibility that enhances the instantaneous motion of an observable (through the factor $e^{2F(0,t)}$ of Sec.~\ref{sec:non-markovian-modification}), may do so in both directions, so that it instead delays the approach of an observable to its pointer value. 
Specifically, when the CP-indivisibility responsible for the enhanced observable current restores a decaying or relaxing quantity, it moves the state away from the pointer value it was approaching. 
The net approach to the pointer value can therefore be delayed rather than accelerated, even as the instantaneous observable current is enhanced.

To make this distinction explicit, let $\mathcal{R}(t)$ denote the deviation of the relaxing observable (the single-channel relaxing quantity of Sec.~\ref{sec:non-markovian-modification}) from its pointer value, i.e., the excited-state population for (zero temperature) amplitude damping or the expectation value of an off-diagonal coherence witness for dephasing. 
It relaxes toward its pointer value as
\begin{equation}
    \mathcal{R}(t) = \mathcal{R}(0)\,e^{-\Gamma(0,t)}, 
    \label{eq:relaxing-quantity}
\end{equation}
with $\Gamma(0,t)$ the integrated canonical decoherence rate. 
We compare the non-Markovian evolution with its matched CP-divisible reference~\eqref{eq:matched-cp-rate} and say the relaxing quantity has reached its pointer value, to within a prescribed threshold $\epsilon$, once $\mathcal{R}(t)\leq\epsilon$. 
Denote the corresponding threshold-crossing times by $t^{\mathrm{M}}_\epsilon$ and $t^{\mathrm{NM}}_\epsilon$, and the delay
\begin{equation}
    \Delta t_\epsilon \equiv t^{\mathrm{NM}}_\epsilon - t^{\mathrm{M}}_\epsilon.
    \label{eq:delay-definition}
\end{equation}
For the matched pair, $\Gamma_{\mathrm{NM}}(0,t)\leq\Gamma_{\mathrm{M}}(0,t)$, so the relaxing quantity remains larger along the non-Markovian solution and the threshold is therefore reached later, giving $\Delta t_\epsilon\geq0$.

At the CP-divisible threshold-crossing time, the non-Markovian solution has accumulated an exponent deficit equal to the difference of the integrated rates~\eqref{eq:Gamma-difference}, $\Gamma_{\mathrm{M}}-\Gamma_{\mathrm{NM}} = 2F(0,t^{\mathrm{M}}_\epsilon)$, which must still be dissipated before the threshold $\epsilon$ is crossed. 
Removing it at a rate no greater than $\gamma_{\max}$, the largest relaxation rate over the interval between the two threshold-crossings, requires
\begin{equation} 
    \Delta t_\epsilon \;\geq\; \frac{2F(0,t^{\mathrm{M}}_\epsilon)}{\gamma_{\max}},
    \qquad
    \gamma_{\max} \equiv \sup_{s\in[t^{\mathrm{M}}_\epsilon,\,t^{\mathrm{NM}}_\epsilon]} \gamma_\mathrm{NM}(s).
    \label{eq:delay-bound}
\end{equation}
The derivation is given in Appendix~\ref{app:delay-bound}. 
When the CP-indivisibility is confined to times before $t^{\mathrm{M}}_\epsilon$ and the relaxation rate has settled to its asymptotic value $\gamma_\infty$, the delay bound~\eqref{eq:delay-bound} becomes independent of the threshold, $\Delta t_\epsilon \geq 2F_\infty/\gamma_\infty$, where $F_\infty = \lim_{t\to \infty} F(0,t) = \int_0^\infty ds\, [-\gamma_\mathrm{NM}(s)]_+ $. 
The bound is then saturated when the residual exponent is removed at the constant maximal rate. 

The accumulated negativity, therefore, carries two distinct operational meanings.
In the driven setting, it multiplies the instantaneous observable speed, 
\begin{equation}
    v_K^{\mathrm{NM}}(t) = e^{2F(0,t)}\,v_K^{\mathrm{M}}(t),
\end{equation}
whereas in the relaxing setting, the same quantity measures a deficit that must subsequently be dissipated, and hence sets a floor on the delay~\eqref{eq:delay-bound}.

\begin{figure}
    \centering
    \includegraphics[width=\columnwidth]{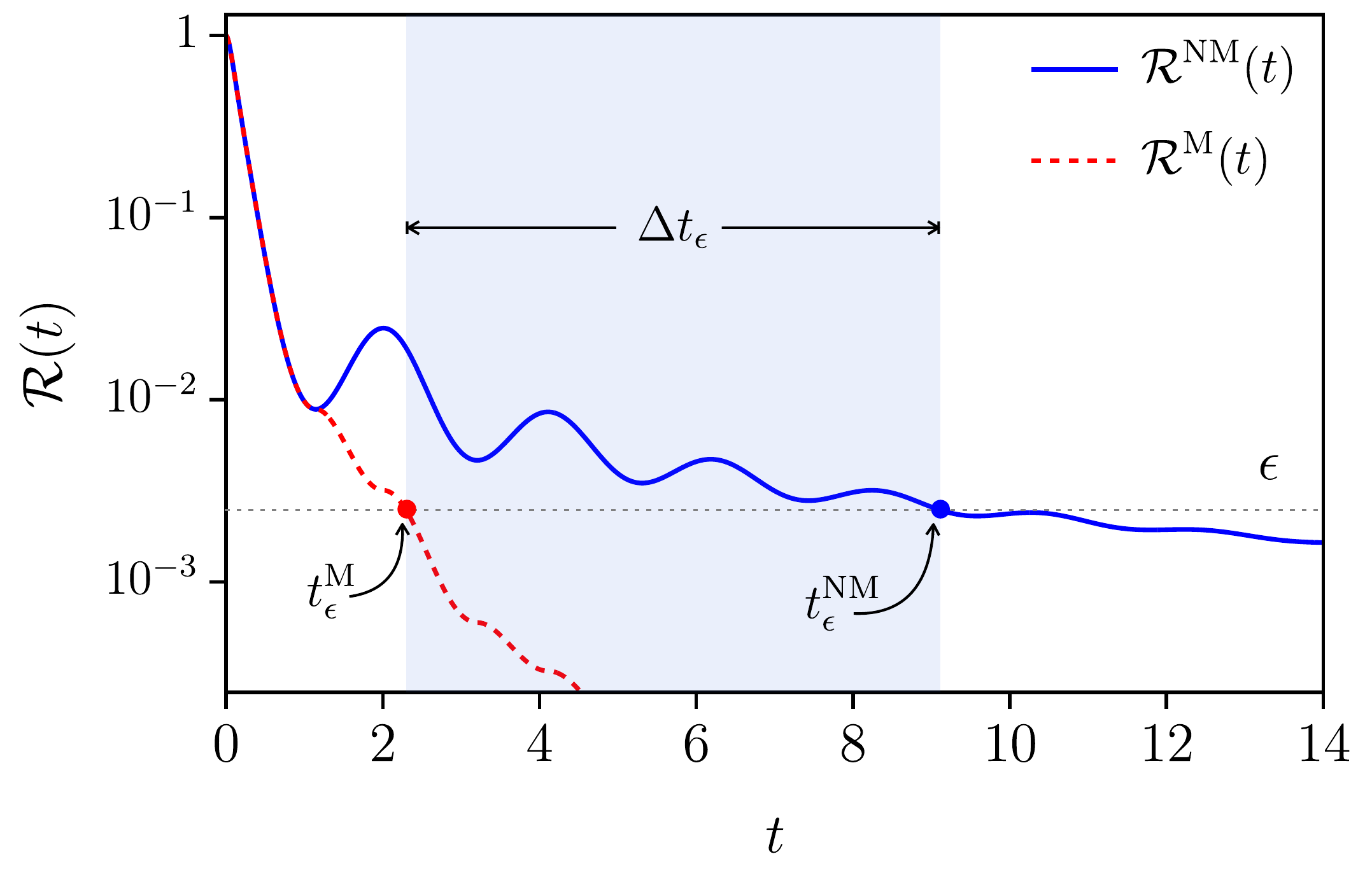}
    \caption{  \label{fig:delay-bound} Non-Markovian delay of einselection for the structured Ohmic-Lorentzian continuum~\eqref{eq:structured-J} with $A=1$, $B=4$, $\omega_0=3$, $\omega_c=10$, and $\lambda=0.3$ (in dimensionless units).  
    The relaxing quantity is the expectation value of the coherence witness $\la \sigma_x \ra$, $\mathcal{R}(t)= 
    e^{-\Gamma(0,t)}$, shown for the non-Markovian evolution (solid blue) and its matched CP-divisible reference (dashed red). 
    The dotted horizontal line indicates the threshold $\epsilon=e^{-\Gamma^*}$, set by $\Gamma^*=6$, and the shaded region corresponds to the delay $\Delta t_\epsilon$~\eqref{eq:delay-definition}.}
  
\end{figure}

Figure~\ref{fig:delay-bound} illustrates the delay~\eqref{eq:delay-definition} for the structured continuum~\eqref{eq:structured-J}. 
The Ohmic background ensures that einselection is permanent, so both threshold-crossing times are finite, while the Lorentzian structure produces the CP-indivisibility that separates them. 
Thus, non-Markovianity can enhance the instantaneous resource available for observable motion while delaying the eventual approach to the pointer value.
The distinction arises because CP-indivisibility restores the relaxing quantity that must subsequently decay before the pointer threshold is reached.

\section{Concluding remarks}
\label{sec:conclusion}

Independent studies had provided apparently contradictory evidence on whether non-Markovian effects hinder or foster the emergence of classical objectivity. 
In the present analysis, we took a step back and focused on the dynamics of einselection, which is the underlying, decohering process required for any quantum system to exhibit objectivity. 
To this end, we examined the dynamical role and influence of non-Markovianity (CP-indivisibility) on quantum speed limits for observables. 
The exact observable motion separates into two contributions, namely a coherent current, bounded by the trace-norm asymmetry between the state and observable, and a dissipative current, through which the environment contributes directly. 
Since the signed currents may reinforce or partially cancel one another, saturation of the speed limit requires both equality in the Schatten-H\"{o}lder bound in addition to compatibility of the signed currents. 

In particular, we identified two mechanisms through which non-Markovian dynamics can modify the speed of an observable.  
Enhancement by non-Markovian (CP-indivisible) dynamics may take place through population restoration (as in amplitude-damping dynamics) or coherence revival (as in pure-dephasing dynamics). Whether einselection is permanent or transient is set by the shape of the environmental spectral density, independently of whether the dynamics is Markovian or non-Markovian. 

A system can rapidly move towards a pointer state but later exhibit a revival, approach the pointer state slowly but remain there permanently, be non-Markovian (CP-indivisible) while still exhibiting permanent decoherence, or have instantaneous speed-limit saturation. 
A negative canonical decoherence rate reverses a signed dissipative current, but it need not increase the absolute speed of an arbitrary observable, particularly when coherent motion or additional dissipative channels are present. 
Moreover, an enhancement of the instantaneous observable speed does not necessarily translate to a larger net observable displacement. 
Ultimately, we find instances in which non-Markovianity can accelerate the instantaneous motion of an observable, while delaying its net einselection, which is characterized through the approach of its expectation value towards its pointer value. 
Concretely, in the single-channel examples, the memory that enhances the instantaneous motion of the relaxing quantity (the expectation value of a coherence witness under dephasing, or the excited-state population under amplitude damping) is the same memory that delays its relaxation. 
Moreover, when einselection is permanent, this delay is bounded below by the accumulated negativity of the canonical decoherence rate divided by the maximal relaxation rate, which we verify in a structured continuum exhibiting both permanent decoherence and CP-indivisibility.

The present framework therefore provides a concrete starting point for distinguishing direct environmental driving, environmentally-induced changes of trace-norm asymmetry, and genuinely history-dependent enhancement of observable motion. 
In conclusion, does non-Markovianity now speed up or slow down Quantum Darwinism? 
Our work has shown that the answer to this question intimately depends on the interplay of the chosen observable, the self-Hamiltonian of the system, and the structure of the environment through which observers access information about a quantum system.

\acknowledgments{S.D. acknowledges support from the John Templeton Foundation under Grant No. 63626. This work was supported by the U.S. Department of Energy, Office of Basic Energy Sciences, Quantum Information Science program in Chemical Sciences, Geosciences, and Biosciences, under Award No. DE-SC0025997. A.B. would like to thank CQT for the support during his visit to NTU.}

\appendix

\section{Schatten–Hölder bound for the coherent current}
\label{app:observable-currents-bounds}

Under the time-local dynamics, the coherent current satisfies
\begin{align}
-i\,\tr{[H(t),\varrho(t)]K} &=-i\,\tr{H(t)[\varrho(t),K]}, 
\end{align}
by linearity and cyclicity of the trace. 
Applying the Schatten–H\"{o}lder inequality,
\begin{equation}
  \left|\tr{A^\dagger B}\right|\leq \norm{A}_p\norm{B}_q, \qquad \frac{1}{p}+\frac{1}{q}=1,
\end{equation}
with $A = H(t),\, B = [\varrho(t),K],\, p = \infty, \,q=1$ yields
\begin{equation} \label{eq:app-schatten-holder}
\left|-i\,\tr{H(t)[\varrho(t),K]}\right| \leq \norm{H(t)}_\infty \norm{[\varrho(t),K]}_1.
\end{equation}
Imposing $\norm{H(t)}_\infty = 1$, 
\begin{equation}
\frac{1}{2} \left|-i\,\tr{[H(t),\varrho(t)]K}\right| \leq \frac{1}{2} \norm{[\varrho(t),K]}_1, 
\end{equation}
where the right-hand side is the definition of the trace-norm asymmetry~\eqref{eq:trace-norm-asymmetry}. 

Equivalently, one may instead choose to impose $\|H(t)\|_\infty\leq 1$. 
For this choice, the H\"{o}lder step~\eqref{eq:app-schatten-holder} (Eq.~\eqref{eq:schatten-holder} in the main text) gives the same bound (since it only weakens the right-hand side of the inequality) and the form of the speed limit is unchanged.
The difference between these two conventions is in their `tightness', where for $\|H\|_\infty=1$ the H\"{o}lder step is tight and $\|[\varrho(t),K]\|_1$ is the largest achievable coherent rate, whereas for $\|H\|_\infty\leq1$ it is a valid upper bound that is tight only when $\|H\|_\infty=1$. 
Note that this choice of normalisation only affects the coherent current. 

It follows that the Schatten-H\"{o}lder inequality~\eqref{eq:app-schatten-holder} is saturated when $H(t)$ is aligned with the sign operator of $-i\,[\varrho(t),K]$.
Define the Hermitian and traceless operator, 
\begin{equation}
    C(t)=-i\,[\varrho(t),K] = \sum_j \lambda_j \ket{c_j}\!\bra{c_j} \quad \lambda_j \in \mathbb{R},
\end{equation} 
where
\begin{equation}
    \norm{[\varrho(t),K]}_1 = \norm{C(t)}_1 = \sum_j \abs{\lambda_j}, 
\end{equation}
so that 
\begin{equation} 
-i\,\tr{H(t)[\varrho(t),K]} =\tr{H(t)C(t)} = \sum_j \lambda_j \bra{c_j} H \ket{c_j}. 
\end{equation}
Recasting the H\"{older} step~\eqref{eq:app-schatten-holder} in terms of $C(t)$,
\begin{align}
\left|\tr{H(t)C(t)} \right| &\leq \norm{H(t)}_\infty \norm{C(t)}_1
\end{align}
For the normalization $\norm{H(t)}_\infty$, the inequality is saturated if 
\begin{equation}
    \abs{\sum_j \lambda_j \bra{c_j}H(t)\ket{c_j}} = \sum_j \abs{\lambda_j}, 
\end{equation}
which holds if each eigenvector $\ket{c_j}$ with $\lambda_j\neq 0$ is an eigenvector of $H(t)$ with the extremal eigenvalue $\lambda_i=\pm\norm{H(t)}_\infty$ (i.e., $\left|\bra{c_j}H\ket{c_j}\right|=\norm{H(t)}_\infty$) \textit{and} all the terms $\lambda_j \bra{c_j}H(t)\ket{c_j}$ have the same sign (so that the modulus of the sum is equal to the sum of the modulus). 
In other words, the inequality is saturated when $H(t)$ is aligned with the sign operator of $C(t)=-i\,[\varrho(t),K]$. 

\section{The adjoint dissipator}
\label{app:adjoint-dissipator}
In this appendix, we analyze the dissipator $\mathcal{D}^\dagger$ in the Heisenberg-picture to elucidate an equivalent, but more useful form. 

\subsection{The adjoint relation}
\label{app:adjoint-relation} 
Let $\mathcal{B}(\mathcal{H})$ denote the vector space of linear operators on the finite-dimensional Hilbert space $\mathcal{H}.$ 
We use the Hilbert-Schmidt inner product $\la A,B\ra _\mathrm{HS} = \tr{B^\dagger A}$. 

For a linear superoperator $\mathcal{D}_t:\mathcal{B}(\mathcal{H})\to \mathcal{B}(\mathcal{H})$, its Hilbert-Schmidt adjoint is defined by
\begin{equation}
    \la \mathcal{D}_t[X],Y \ra _\mathrm{HS} = \la X, \mathcal{D}_t^\dagger [Y]\ra _\mathrm{HS}
\end{equation} for all $X,Y \in \mathcal{B(\mathcal{H)}}$. 
By definition of the Hilbert-Schmidt inner product, 
\begin{equation}
    \tr{Y^\dagger\mathcal{D}_t[X]}= \tr{(\mathcal{D}_t^\dagger[Y])^\dagger X}. 
\end{equation}
Taking $X = \varrho (t), Y=K,$ with $K=K^\dagger$, 
\begin{align}
    \tr{\mathcal{D}_t[\varrho(t)]K} = \tr{(\mathcal{D}_t^\dagger[K])^\dagger\varrho(t)} = \la(\mathcal{D}_t^\dagger[K])^\dagger\ra_t. 
\end{align}

We now find the form of $(\mathcal{D}_t^\dagger[K])^\dagger$.
Let $\varrho(t)$ be a density operator on $\mathcal{H}$, and $K=K^\dagger$ and $L_\mu(t)$ be bounded operators on $\mathcal{H}$ such that $\tr{L_\mu(t)}=0, \, \tr{L_\mu^\dagger(t)L_\nu(t)}=\delta_{\mu\nu}.$

For brevity, we suppress the explicit time dependence of operators and constants. 

\begin{align}
(\mathcal{D}_t^\dagger[K])^\dagger  &= \left( \sum_\mu \gamma_\mu \left( L_\mu^\dagger K L_\mu -  \frac{1}{2} \left\{ L_\mu^\dagger L_\mu,K \right\} \right) \right)^\dagger \nonumber \\
&= \sum_\mu \gamma_\mu^\dagger \left( \left(L_\mu^\dagger K L_\mu\right)^\dagger -  \frac{1}{2} \left\{ L_\mu^\dagger L_\mu,K \right\}^\dagger \right) \nonumber \\
&= \sum_\mu \gamma_\mu(L_\mu^\dagger KL_\mu - \frac{1}{2}\{L_\mu^\dagger L_\mu,K\}) = \mathcal{D}_t^\dagger[K], 
\end{align}
where we have used the Hermiticity of $K$. 

Since $t$ was chosen arbitrarily, this identity holds throughout the time interval on which the generator is defined.

\subsection{Commutator representation}
\label{app:adjoint-commutator-form}

Recall that the adjoint dissipator acting on the observable $K$ is given by Eq.~\eqref{eq:adjoint-dissipator}. Fix $\mu$ and $t$, and temporarily suppress the explicit time dependence, 
\begin{align}
\mathcal{D}^\dagger[K]
&= \frac{1}{2} \sum_\mu \gamma_\mu (2L_\mu^\dagger KL_\mu - \left(L_\mu^\dagger L_\mu K + K L_\mu^\dagger L_\mu \right))  \nonumber \\
&= \frac{1}{2} \sum_\mu \gamma_\mu (L_\mu^\dagger \left(KL_\mu - L_\mu K \right) + \left(L_\mu^\dagger K - K L_\mu^\dagger\right) L_\mu ) \nonumber \\
&= \frac{1}{2} \sum_\mu \gamma_\mu (L_\mu^\dagger \left[K,L_\mu\right]+ \left[L_\mu^\dagger,K \right] L_\mu ). 
\end{align}

Restoring the explicit time dependence yields the commutator representation
\begin{align} \label{eq:app-commutator}
    \mathcal{D}_t^\dagger[K] = \frac{1}{2} \sum_\mu &\gamma_\mu(t) \left(  L_\mu^\dagger(t)[K,L_\mu(t)] + [L_\mu^\dagger(t),K]L_\mu(t) \right).
\end{align}
In the case that each jump operator is Hermitian, $L_\mu(t)=L_\mu^\dagger(t)$, this now takes the form of a double commutator, 
\begin{align} \label{eq:double-commutator-hermitian}
    \mathcal{D}_t^\dagger[K] &= \frac{1}{2} \sum_\mu \gamma_\mu(t) \left(  L_\mu(t)[K,L_\mu(t)] 
    + [L_\mu(t),K]L_\mu(t) \right) \nonumber \\ 
    &= \frac{1}{2} \sum_\mu \gamma_\mu(t) \left(  -L_\mu(t)[L_\mu(t),K] 
    + [L_\mu(t),K]L_\mu(t) \right) \nonumber \\ 
    &= -\frac{1}{2}   \sum_\mu \gamma_\mu(t)
    \left[L_\mu(t), [L_\mu(t),K] \right].
\end{align}

\subsection{Interpretation via trace-norm asymmetry}
\label{app:adjoint-commutator-form-asymmetry}
By taking the absolute expectation value of the left-hand side of Eq.~\eqref{eq:app-commutator} (Eq.~\eqref{eq:commutator-dissipator} in the main text) and using the H\"{o}lder inequality, we obtain a bound for the action of the adjoint dissipator on an observable in terms of the trace-norm noncommutativity between $L_{\mu}$ and $K$. 
Temporarily suppress the explicit time dependence,  
\begin{align}
    &\left|\la \mathcal{D}_t^\dagger[K] \ra \right| = 
    \left| \frac{1}{2} \sum_\mu \gamma_\mu \la \left( L_\mu^\dagger[K,L_\mu] + [L_\mu^\dagger,K]L_\mu \right) \ra \right| \nonumber \\
    &\leq \frac{1}{2} \sum_\mu \left|\gamma_\mu\right| \left(
    \left|\la L_\mu^\dagger [K,L_\mu] \ra \right| +\left|\la[L_\mu^\dagger,K] L_\mu(t) \ra\right| \right)  \nonumber
\end{align}
where
\begin{equation}
    \left|\la B \ra\right| = \left|\tr{\varrho B}\right| \leq \norm{\varrho}_\infty \norm{B}_1. 
\end{equation}
The operator norm (or spectral norm) of a density matrix $\varrho$ is equal to its largest eigenvalue, $0 \leq \norm{\varrho(t)}_\infty \leq 1$.  Namely, the operator norm/H\"{o}lder step is submultiplicative,
\begin{align}
    \left|\la L_\mu^\dagger [K,L_\mu] \ra \right| &\leq \norm{L_\mu}_1 \norm{[K,L_\mu]}_1 \\
    \left|\la [L_\mu^\dagger,K] L_\mu \ra \right| &\leq \norm{[L_\mu^\dagger,K]}_1 \norm{L_\mu}_1, 
\end{align} 
where we have also used $\norm{L_\mu^\dagger}_1 = \norm{L_\mu}_1$. 
Furthermore, since $K$ is an observable (and therefore Hermitian), $[K,L_\mu]^\dagger = [L_\mu^\dagger, K]$, and the trace-norm noncommutativity (and therefore the \textit{asymmetry}) between $L_{\mu}(t)$ and $K$ is equivalent to that between $L_{\mu}^\dagger(t)$ and $K$, 
\begin{equation}
    \norm{[K,L_\mu]}_1 = \norm{[L_\mu^\dagger,K]}_1. 
\end{equation}
Altogether, restoring time dependence, 
\begin{align}
    \left|\la \mathcal{D}_t^\dagger[K] \ra \right|
    &\leq \sum_\mu \left|\gamma_\mu(t)\right| \norm{L_\mu(t)}_1 \norm{[K,L_\mu(t)]}_1. 
\end{align}
which can be understood from the viewpoint of asymmetry of the decoherence operators with respect to the observable $K$ as a resource for quantum speed. If we set the jump operators to be Pauli (ladder) operators, $\norm{L_\mu}_1=1$.

\section{Exactly solvable dynamical models}
\label{app:exact-dynamical-models}
In this appendix, we consider the exactly solvable dynamical models in detail.

\subsection{Time-homogeneous amplitude-damping semigroup}
\label{app:amplitude-damping}
Consider a qubit undergoing amplitude damping at a constant rate, as
introduced in Sec.~\ref{sec:instantaneous-bound}, whose Hilbert space is $\mathcal{H}=\mathbb{C}^2$. 
We choose a single decoherence operator $L=\sigma_-=\ket{0}\!\bra{1}$, which is Hilbert-Schmidt normalized $\tr{L^\dagger L} = \tr{\sigma_+\sigma_-} = 1$.
We take the initially excited state $\varrho(0)=\ket{1}\!\bra{1}$, the observable $K=\sigma_z$, and a unit-norm Hamiltonian $\norm{H}_\infty=1$ that commutes with $\sigma_z$, e.g., $H=\openone$ or $H=\sigma_z$.  
Because $\varrho(0)$ is diagonal and $[H,\sigma_z]=0$, the state remains diagonal for all $t$; a noncommuting $H$ would instead rotate population into coherence. 
With $\varrho(t)$ and $K=\sigma_z$ then simultaneously diagonal, $[\varrho(t),K]=0$, so the coherent current $-i\,\tr{H[\varrho(t),K]}$ vanishes and the observable motion is generated entirely by the dissipative current.

Note that once the state is diagonal the coherent current vanishes for any $H$, since it depends only on $[\varrho(t),K]$; the commutation condition on $H$ serves only to keep the solution diagonal in the first place.
In the language of the later sections, $\sigma_z$ here is a coherence witness driven directly by the dissipator, $\mathcal{D}_t^\dagger[\sigma_z]\neq0$, whose expectation value the environment carries to its stationary value through the dissipative current alone. 
(This is in contrast to the pure-dephasing channel treated in Appendix~\ref{app:pure-dephasing}, where $\sigma_z$ is instead the pointer observable left invariant by the dissipator.)
For a constant canonical decoherence rate $\gamma>0$, the master equation is given by Eq.~\eqref{eq:amplitude-damping} in the main text, here reproduced
\begin{equation}
\begin{aligned} \label{eq:app-amplitude-damping}
    \dot{\varrho}(t) &= - i [\sigma_z,\varrho(t)] + \gamma \left(\sigma_-\varrho(t)\sigma_+ - \frac{1}{2} \left\{ \sigma_+\sigma_-, \varrho(t) \right\} \right).
\end{aligned}
\end{equation}
Writing the state in the computational basis, 
\begin{equation}
     \varrho(t) = \begin{pmatrix}
        \varrho_{00}(t) & \varrho_{01}(t)\\
        \varrho_{10}(t) & \varrho_{11}(t)
    \end{pmatrix},
\end{equation}
the populations and coherences evolve according to 
\begin{align}
    \dot{\varrho}_{00}(t)&=\gamma\varrho_{11}(t),\\
    \dot{\varrho}_{11}(t)&=-\gamma\varrho_{11}(t),\\
    \dot{\varrho}_{01}(t)&=-\left(\frac{\gamma}{2}+2 i\right)\varrho_{01}(t),\\
    \dot{\varrho}_{10}(t)&=-\left(\frac{\gamma}{2}-2 i\right)\varrho_{10}(t).
\end{align}
For the initially excited state $\varrho(0)=\ket{1}\!\bra{1}$, the initial coherences vanish. 
The coherence equations therefore imply $\varrho_{01}(t)=\varrho_{10}(t)=0$ for all $t\geq0$, while $\varrho_{11}(t)=e^{-\gamma t}$.
Trace preservation then gives Eq.~\eqref{eq:amplitude-damping-solution},  
\begin{equation}
    \varrho(t) = \begin{pmatrix}
        1-e^{-\gamma t} & 0\\
        0 & e^{-\gamma t}
    \end{pmatrix}.
\end{equation}
For the choice of observable $K=\sigma_z$, since both $\varrho(t)$ and $K$ are diagonal in the $\sigma_z$ eigenbasis, $[\varrho(t),K]=0$ for all $t\geq 0$. 
Consequently, the trace-norm asymmetry vanishes throughout the evolution, $\frac{1}{2} \left\|[\varrho(t),K] \right\|_1 = 0$, and the coherent current vanishes identically $J_K^\mathrm{coherent}(t)=0$. 
The expectation value of $K$ is
\begin{align}
    \la K\ra_t = \tr{\varrho(t)\sigma_z} = 1-2e^{-\gamma t}, 
\end{align}
and the instantaneous observable speed~\eqref{eq:speed-definition} is then 
\begin{equation}
    v_K(t) = \frac{1}{2} \left| \frac{d}{dt}\la K\ra_t \right| = \gamma e^{-\gamma t}.
\end{equation}
We now turn to computing the instantaneous obervable speed limit~\eqref{eq:instantaneous-bound}. 
The dissipator is time independent, $\mathcal{D}_t\equiv\mathcal{D}$, and its adjoint acts on $K=\sigma_z$ according to
\begin{align}
    \mathcal{D}^\dagger[\sigma_z]
    &= \gamma \left( \sigma_+\sigma_z\sigma_- -
        \frac{1}{2} \left\{ \sigma_+\sigma_-, \sigma_z \right\} \right) \nonumber \\
    &=  2\gamma\ket{1}\!\bra{1} \nonumber \\
    &= \gamma \left( \openone-\sigma_z \right).
\end{align}
Therefore, the dissipative current $J_K^\mathrm{dissipative}(t)$ is then
\begin{align}
    \la \mathcal{D}^\dagger[\sigma_z] \ra_t
    &= 2\gamma \,\tr{\varrho(t)\ket{1}\!\bra{1}} 
    = 2\gamma e^{-\gamma t}, 
\end{align}
and the instantaneous observable-speed bound~\eqref{eq:instantaneous-bound} reduces to
\begin{align}
    \frac{1}{2} \left| J_K^\mathrm{dissipative} (t) \right| = \frac{1}{2} \left( \left| \la \mathcal{D}^\dagger[K] \ra_t \right| \right) = \gamma e^{-\gamma t}.
\end{align}
Comparison with the exact speed shows that the instantaneous bound is saturated for all $t\geq0$,
\begin{equation}
    v_K(t) = \frac{1}{2} \left| \la \mathcal{D}^\dagger[K] \ra_t \right|.
\end{equation}

\subsection{Time-homogeneous pure-dephasing semigroup}
\label{app:pure-dephasing}
We consider the time-homogeneous pure-dephasing dynamics introduced in Sec.~\ref{sec:dissipative-currents-pointer-observables}.
The system is a qubit with Hilbert space $\mathcal{H}=\mathbb{C}^2$. 
We choose the unit-norm Hamiltonian $H=\sigma_z$, and a single Hermitian decoherence operator $L=\frac{\sigma_z}{\sqrt{2}}$ which is Hilbert--Schmidt normalized since $\tr{L^\dagger L} = \frac{1}{2}\tr{\sigma_z^2} = 1$.
For a constant dephasing rate $\gamma>0$, the master equation is given by Eq.~\eqref{eq:pure-dephasing} in the main text, 
\begin{equation} \label{eq:app-pure-dephasing}
    \dot{\varrho}(t)= -i[\sigma_z,\varrho(t)] + \frac{\gamma}{2} \left( \sigma_z\varrho(t)\sigma_z - \varrho(t) \right).
\end{equation}
We take the initial state $\varrho(0) = \ket{+}\!\bra{+}$ and choose the observable $K=\sigma_x$.
Writing the state in the computational basis, the master equation gives
\begin{align}
    \dot{\varrho}_{00}(t) &= 0, \\
    \dot{\varrho}_{11}(t) &= 0, \\
    \dot{\varrho}_{01}(t) &= -\left( \gamma+2 i \right) \varrho_{01}(t), \\
    \dot{\varrho}_{10}(t) &= -\left( \gamma-2 i \right) \varrho_{10}(t). 
\end{align}
The initial state satisfies
\begin{equation}
    \varrho_{00}(0) = \varrho_{11}(0) = \varrho_{01}(0) = \varrho_{10}(0) = \frac{1}{2}, 
\end{equation}
so the populations are constant, while the coherences decay. 
Hence,
\begin{equation}
    \varrho(t) = \frac{1}{2} \begin{pmatrix}
        1 & e^{-(\gamma+2 i )t} \\
        e^{-(\gamma-2 i)t} & 1
    \end{pmatrix}.
\end{equation}
Equivalently, in Bloch-vector form,
\begin{equation} \label{eq:pure-dephasing-bloch}
    \varrho(t) = \frac{1}{2} \left[ \openone + e^{-\gamma t} \cos{(2t)}\sigma_x + e^{-\gamma t} \sin{(2t)}\sigma_y \right].
\end{equation}
The expectation value of $K=\sigma_x$ is
\begin{equation}
    \la K\ra_t = \tr{\varrho(t)\sigma_x} = e^{-\gamma t}\cos{(2t)}, 
\end{equation}
whose time derivative is
\begin{equation}
    \frac{d}{dt}\la K\ra_t = -e^{-\gamma t} \left[ \gamma\cos{(2t)} + 2\sin{(2t)} \right]. 
\end{equation}
The instantaneous observable speed is therefore
\begin{equation} \label{eq:pure-dephasing-instantaneous-speed}
    v_K(t) = \frac{e^{-\gamma t}}{2} \left|\gamma\cos{(2t)} + 2\sin{(2t)}\right|, 
\end{equation}
which can also be attained using Eq.~\eqref{eq:instantaneous-speed-currents} via the coherent and dissipative currents. 

Now, consider the instantaneous observable speed limit~\eqref{eq:instantaneous-bound} for which we will explicitly compute the currents. 
We first evaluate the coherent current.
Using Eq.~\eqref{eq:pure-dephasing-bloch}, 
\begin{align}
    [\varrho(t),\sigma_x] &= \frac{e^{-\gamma t}}{2}  \sin{(2t)} [\sigma_y,\sigma_x] = - i e^{-\gamma t} \sin{(2t)}\sigma_z.
\end{align}
Since the singular values of $\sigma_z$ are both equal to one,
\begin{equation}
    \norm{[\varrho(t),\sigma_x]}_1 = 2e^{-\gamma t} \left|\sin{(2t)}\right|,
\end{equation}
and the corresponding coherent current $J_K^\mathrm{coherent}(t)$~\eqref{eq:coherent-current} is 
\begin{align}
    - i\,\tr{H[\varrho(t),K]} &= -i\,\tr{\sigma_z[\varrho(t),\sigma_x]} \nonumber \\
    &= -2e^{-\gamma t}\sin{(2t)}.
\end{align}

The dissipator is time independent, so we write 
$\mathcal{D}_t\equiv\mathcal{D}$. 
Since $L=L^\dagger$, the adjoint dissipator action on $K$ may be evaluated using the double-commutator representation~\eqref{eq:double-commutator-hermitian}. 
For $L=\sigma_z/\sqrt{2}$ and $K=\sigma_x$,
\begin{align}
    \mathcal{D}^\dagger[K]=-\frac{\gamma}{2}[L,[L,\sigma_x]]
    &= -\frac{\gamma}{4} [\sigma_z,[\sigma_z,\sigma_x]] = -\gamma \sigma_x, 
\end{align}
whose expectation value along the solution is the dissipative current $J_K^\mathrm{dissipative}(t)$~\eqref{eq:dissipative-current}, 
\begin{equation}
    \la \mathcal{D}^\dagger[\sigma_x] \ra_t = -\gamma e^{-\gamma t}\cos{(2t)}. 
\end{equation}
Altogether, the instantaneous observable speed limit~\eqref{eq:instantaneous-bound} is 
\begin{equation} \label{eq:pure-dephasing-instantaneous-bound}
    v_K(t) \leq e^{-\gamma t} \left[ \left|\sin{(2t)} \right|+ \frac{\gamma}{2} \left|\cos{(2t)}\right| \right].
\end{equation}

The difference between the exact speed~\eqref{eq:pure-dephasing-instantaneous-speed} and this upper bound~\eqref{eq:pure-dephasing-instantaneous-bound} is determined by the relative signs of the coherent and dissipative contributions before the triangle inequality is applied.

\subsection{Microscopic models and spectral conditions}
\label{app:microscopic-models}
In this section, we keep the same observable setup but replace the phenomenological decoherence rates with those from microscopic models, so that the non-Markovianity is generated microscopically by a bath with some spectral density. 

\subsubsection{Damped Jaynes-Cummings}
\label{app:damped-jaynes-cummings}
For completeness, we first consider the exact decoherence rate of the damped Jaynes-Cummings model~\cite{Garraway1997, Breuer1999}, so that the non-Markovianity is generated microscopically by a Lorentzian bath. 

We work in the interaction picture, with $H=\openone$, $L=\sigma_-$, $K=\sigma_z$, and the initially excited two-level system $\varrho(0)=\ket{1}\!\bra{1}$ so $\varrho_{11}(0)=1$. 
A two-level system resonantly coupled to a leaky single-mode cavity (that is, an initially vacuum bath described by a Lorentzian spectral density of width $\lambda$ and coupling $\gamma_0$) has the time-local amplitude-damping generator with canonical decoherence rate (e.g.,~\cite{Deffner2013})
\begin{align}
  \gamma(t)&=\frac{2\gamma_0\lambda\,\sinh{(dt/2)}}
                {d\cosh{(dt/2)}+\lambda\sinh{(dt/2)}}, \\      
   d&=\sqrt{\lambda^2-2\gamma_0\lambda}.
\end{align}
As in Appendix~\ref{app:amplitude-damping}, the state stays diagonal,
\begin{align}
  \varrho(t)&=\begin{pmatrix} 1-e^{-\Gamma(t)}&0\\0&e^{-\Gamma(t)} \end{pmatrix},\\
  \Gamma(t)&=\int_0^t ds\,\gamma(s),
\end{align}
with the excited-state population
\begin{equation}
  \varrho_{11}(t)=e^{-\Gamma(t)}
    =e^{-\lambda t}\left|\cosh{\frac{dt}{2}}+\frac{\lambda}{d}\sinh{\frac{dt}{2}}\right|^2. 
\end{equation}
The coherent current vanishes identically, and the motion is purely dissipative,
\begin{equation}
  \la \mathcal{D}^\dagger_t[\sigma_z]\ra_t
    =2\gamma(t)\,\varrho_{11}(t)=2\gamma(t) e^{-\Gamma(t)}.
\end{equation}
With $\la\sigma_z\ra_t=1-2e^{-\Gamma(t)}$,
\begin{align}
  \frac{d}{dt}\la\sigma_z\ra_t &= 2\gamma(t) e^{-\Gamma(t)}, \\
  v_K(t)&=\frac{1}{2} \left|\frac{d}{dt}\la\sigma_z\ra_t \right|
        =|\gamma(t)|\,e^{-\Gamma(t)},
\end{align}
so the instantaneous bound~\eqref{eq:instantaneous-bound} is saturated for all $t\geq0$, 
\begin{equation}
  v_K(t)\le\frac{1}{2}\left(\|[\varrho,\sigma_z]\|_1
        +|\la \mathcal{D}^\dagger_t[\sigma_z]\ra_t|\right)
        =|\gamma(t)|\,e^{-\Gamma(t)}.
\end{equation}
At a population zero, $\gamma(t)$ diverges, while the observable speed $v_K(t)$ stays finite (derivative of a smooth function) and vanishes at the turning point, i.e., the observable speed is regular where the canonical decoherence rate is singular. 

\paragraph*{Weak coupling limit (Markovian dynamics).}
For $\gamma_0<\lambda/2$, $d$ is real and $\gamma(t)>0$ for all $t$, so that $\la\sigma_z\ra$ increases monotonically from $-1$ to $+1$, the rate $2\gamma(t) e^{-\Gamma(t)}$ is sign-definite. 

\paragraph*{Strong coupling limit (non-Markovian dynamics).}
In the non-Markovian regime $\gamma_0>\lambda/2$, $d$ becomes imaginary. 
Writing $d=i\tilde{d}$ with $\tilde{d}=\sqrt{2\gamma_0\lambda-\lambda^2}$, the population takes the form
\begin{equation}
  \varrho_{11}(t)=e^{-\lambda t}\left[\cos{\frac{\tilde{d} t}{2}}
    +\frac{\lambda}{\tilde{d}}\sin{\frac{\tilde{d} t}{2}}\right]^2, 
\end{equation}
where the hyperbolic functions become trigonometric, and the argument of the modulus is real, so the modulus squared is the argument squared. 
The first revival, where the excited-state population momentarily vanishes, occurs when $\varrho_{11}(t)=0$. 
Since $e^{-\lambda t}>0$ for all $t$, this requires the bracketed factor to vanish,
\begin{align}
  \tan{\frac{\tilde{d} t}{2}}&=-\frac{\tilde{d}}{\lambda}.
\end{align}
Solving for the first positive root (since $\tilde{d}/\lambda>0$, the tangent must be negative, which \textit{first} occurs in the second quadrant, so
$\tilde{d} t/2=\pi-\arctan(\tilde{d}/\lambda)$), gives
\begin{equation}
  t_{\mathrm{revival}}=\frac{2}{\tilde{d}}
    \left(\pi-\arctan\frac{\tilde{d}}{\lambda}\right).
\end{equation}

\subsubsection{Jaynes-Cummings with pure dephasing}

The Jaynes-Cummings paradigmatic setting is a two-level system (qubit) of frequency $\omega_q$ coupled to a single bosonic cavity mode of frequency $\omega_c$ at detuning $\Delta=\omega_q-\omega_c$. 
In the rotating-wave approximation, which drops the counter-rotating terms $\sigma_+ a^\dagger+\sigma_- a$, the coupling reduces to the Jaynes-Cummings form,
\begin{equation}
  H_\mathrm{JC}=\frac{1}{2}\omega_q\,\sigma_z+\omega_c\,a^\dagger a
    +g\left(\sigma_+ a+\sigma_- a^\dagger\right),
\end{equation}
where the coupling term, of strength $g$, exchanges excitations between the qubit and cavity.
On resonance this term drives amplitude damping (Appendix~\ref{app:damped-jaynes-cummings}), but we now want to explore the opposite regime. 
When the detuning is large compared with the coupling, $|\Delta|\gg g$, the exchange is off-resonant and cannot transfer a real excitation. 

Technically, a \textit{Schrieffer-Wolff} transformation eliminates the coupling term to second order in $g/\Delta$ and replaces it with an effective dispersive interaction~\cite{LarsonMavrogordatos2021, Carbonaro1979, Klimov2002}
\begin{align}
  H_{\mathrm{dispersive}}&=\frac{1}{2}\,(\omega_q+\chi)\,\sigma_z
    +\left(\omega_c+\chi\,\sigma_z\right)a^\dagger a, 
\end{align}
where $\chi=g^2/\Delta$ and the interaction $\chi\,\sigma_z\,a^\dagger a$ is diagonal in $\sigma_z$.
From the qubit perspective, this commutes with $\sigma_z$, so $[H_{\mathrm{dispersive}},\sigma_z]=0$ and the populations in the $\sigma_z$ basis do not change. 
From the cavity perspective, $((\omega_c+\chi \sigma_z) a^\dag a)$ shifts the resonance to $\omega_c\pm\chi$ conditioned on the qubit state, so the field evolves into two distinguishable configurations depending on the qubit state and the coherence decays. 
Therefore, this describes pure dephasing, since the effective jump operator $L\propto\sigma_z$ is Hermitian (hence normal) and the pointer basis is $\{\ket0,\ket1\}$.

We again work in the interaction picture with respect to the qubit free Hamiltonian and the bare cavity Hamiltonian $\omega_c a^\dagger a$, leaving the dispersive interaction $\chi\sigma_z a^\dagger a$. 
In order to obtain nontrivial open-system dephasing dynamics, we include cavity damping at rate $\kappa$, described by the Lindblad term $\kappa \mathcal{D}[a]$. 
To understand why, first consider the case without loss ($\kappa=0$). 
Start with the qubit in a superposition and the cavity in a coherent state $\ket{\beta_0}$,
\begin{equation}
  \ket{\psi(0)}=(c_0\ket0+c_1\ket1)\otimes\ket{\beta_0},
\end{equation}
Since $\chi\sigma_z a^\dagger a$ is diagonal in the qubit basis, the qubit label is conserved and the two branches of the initial state $\ket{\psi(0)}$ evolve independently. 

Conditioned on $\sigma_z = z = \pm1$, the cavity sees an effective oscillator $H^{(z)}_\mathrm{cavity}=\chi z a^\dagger a$, under which a coherent state stays coherent with its amplitude rotating $\ket{\beta_0}\to \ket{\beta_0 e^{-i\chi zt}}$. 
The two conditional cavity states counter-rotate in phase space without decaying,
\begin{align}
  \alpha_{0}(t) &= \beta_0\, e^{- i\chi t}, \\
  \alpha_{1}(t) &= \beta_0\, e^{+ i\chi t},
\end{align}
with 
\begin{equation} \label{eq:amplitude-branches}
    |\alpha_0(t)|^2=|\alpha_1(t)|^2=\overline{n}, 
\end{equation}
where $\overline{n} = \langle n(0)\rangle = |\beta_0|^2$ is the (initial) mean photon number (conserved for $\kappa=0$). 
With no loss, the amplitudes only acquire a phase, so only the relative phase between the branches evolves.

The joint (pure) state at time $t \geq 0$ is 
\begin{equation}
  \ket{\psi(t)} = c_0\ket{0}\otimes\ket{\alpha_0(t)} 
  + c_1\ket{1}\otimes\ket{\alpha_1(t)}.
\end{equation}
Forming $\ket{\psi(t)}\!\bra{\psi(t)}$ and tracing out the cavity leaves the overlap of the conditional cavity fields on the qubit's off-diagonal element, i.e., the $\ket{0}\!\bra{1}$ term picks up the factor $\tr{\ket{\alpha_0(t)}\!\bra{\alpha_1(t)}}=\braket{\alpha_1(t)}{\alpha_0(t)}$, so that
\begin{align}
  \varrho_{01}(t) &= c_0 c_1^*\,\braket{\alpha_1(t)}{\alpha_0(t)} \nonumber \\
              &= \varrho_{01}(0)\,\braket{\alpha_1(t)}{\alpha_0(t)},
\end{align}
where we have used $\varrho_{01}(0)=c_0c_1^*$ and $\braket{\alpha_1(0)}{\alpha_0(0)}=1$. 
The time dependence of the coherence is the overlap of the conditional cavity states, i.e., the more distinguishable the field becomes for the qubit states (the smaller the overlap), the more the coherence is suppressed.

The overlap of two coherent states is
\begin{align}
    \braket{\alpha_1(t)}{\alpha_0(t)} =e^{-\frac{1}{2}|\alpha_0(t)|^2-\frac{1}{2}|\alpha_1(t)|^2
      +\alpha_1^*(t)\alpha_0(t)}.
\end{align}
With $\kappa=0$, both branches have equal, time-independent magnitude given by Eq.~\eqref{eq:amplitude-branches}, so the first two terms together give $-\overline{n}$, while the cross term gives
\begin{equation}
  \alpha_1^*(t)\alpha_0(t)
    =\left(\beta_0 e^{+i\chi t}\right)^*\left(\beta_0 e^{-i\chi t}\right)
    =\overline{n}\,e^{-2i\chi t}.
\end{equation}
Altogether, the overlap of two coherent states becomes 
\begin{equation}
  \braket{\alpha_1(t)}{\alpha_0(t)}
    =\exp{-\overline{n}\left(1-e^{-2i\chi t}\right)},
\end{equation}
whose absolute value is $\left|\braket{\alpha_1(t)}{\alpha_0(t)}\right|=e^{-\Lambda(t)}$, with 
\begin{equation}
  \Lambda(t) = \overline{n}\left(1-\cos{2\chi t}\right) \in [0,\,2\overline{n}],
\end{equation}
which returns to zero when $2\chi t = 2\pi m$, $m\in\mathbb{Z}^+$. 
The qubit coherence
\begin{equation}
    |\varrho_{01}(t)| = |\varrho_{01}(0)|\,e^{-\Lambda(t)}
\end{equation}
therefore dips and fully revives periodically, so it is only temporarily stored in qubit-cavity entanglement that periodically disentangles.
While the joint qubit-cavity dynamics are closed, unitary, recurrent evolution, (i.e., not `decoherence'), the reduced qubit dynamics are nonunitary but reversible. 
The revivals here (in the $\kappa=0$ dispersive limit) are due to reversible qubit–cavity entanglement in a closed system.

\subsubsection{Jaynes-Cummings with dephasing and damping}

Including cavity damping at rate $\kappa$ makes the cavity an open subsystem whose conditional fields decay toward the vacuum while correlations are also carried into the output field. 
The conditional cavity amplitudes acquire the envelope $e^{-\kappa t/2}$, so the overlap of the fields gives the modified form
\begin{equation}
  \tilde{\Lambda}(t) = \overline{n}\, e^{-\kappa t}\left(1-\cos{2\chi t}\right). 
\end{equation}
Again, the dispersive coupling $\chi$ continually creates qubit-dependent cavity fields, but now damping removes photons from the cavity and transfers correlations to the external reservoir. 
This competition between the dispersive evolution and cavity leakage sets the memory timescale of the reduced qubit dynamics. 

The coherence is
\begin{equation}
  \varrho_{01}(t)=\varrho_{01}(0)\,
    \exp{-\overline{n}\,e^{-\kappa t}\left(1-e^{-2i\chi t}\right)}.
\end{equation}
Splitting 
\begin{equation}
    1-e^{-2i\chi t}=(1-\cos{2\chi t})+i\sin{2\chi t}
\end{equation} separates the exponent into a real dephasing part and an imaginary bath-induced phase,
\begin{equation}\varrho_{01}(t)=\varrho_{01}(0)\,e^{-\Gamma(t)}e^{-i\Phi(t)},
\end{equation}
with
\begin{align}
  \Gamma(t)&=\overline{n}\,e^{-\kappa t}\left(1-\cos{2\chi t}\right),\\
  \Phi(t)&=\overline{n}\,e^{-\kappa t}\sin{2\chi t}, 
\end{align}
where $\Gamma(t)\ge0$ is the decoherence function anticipated above (the lossy $\tilde\Lambda(t)$), which suppresses $|\varrho_{01}|$, and $\Phi(t)$ is the bath-induced phase (i.e., a coherent $\sigma_z$-rotation). 
The closed-cavity limit $\kappa\to0$ recovers the reversible lossless case $\Gamma(t)\to\Lambda(t)$. 

Since $\chi\sigma_z a^\dagger a$ commutes with $\sigma_z$ and the loss acts only on the cavity field, the populations remain unchanged and the dynamics is dephasing.
The canonical dephasing rate is
\begin{align}
  \gamma(t)&= \dot\Gamma(t) \nonumber \\
  &=\overline{n}\,e^{-\kappa t}\left[2\chi\sin{2\chi t}-\kappa(1-\cos{2\chi t})\right], 
\end{align}
where $2\chi\sin{2\chi t}$ describes the coherent dispersive evolution that periodically builds and reduces the distinguishability of the conditional cavity fields, while $-\kappa(1-\cos{2\chi t})$ arises from cavity leakage, which transfers correlations to the external reservoir. 

The rate is transiently negative when
\begin{align}
  2\chi\sin{2\chi t} &<\kappa(1-\cos{2\chi t}) \\
  2\chi\left(2\sin{\chi t}\cos{\chi t}\right)&<\kappa\left(2\sin^2{\chi t}\right). 
\end{align}
Dividing by $2\sin{\chi} t>0$ on the first interval $0<\chi t<\pi$,
\begin{align}
  2\chi\cos{\chi t} &<\kappa\sin{\chi t} \\
  \tan{\chi t} &> \frac{2\chi}{\kappa}, 
\end{align}
and the \textit{first} backflow therefore begins at
\begin{equation}
  t_{\mathrm{backflow}}=\frac{1}{\chi}\arctan{\frac{2\chi}{\kappa}}.
\end{equation}
In the closed-cavity limit $\kappa\to0$, $\arctan{(2\chi/\kappa)}\to\pi/2$ and $t_{\mathrm{backflow}}\to\pi/2\chi$. 
In the heavily-damped limit $\kappa\to\infty$, $t_{\mathrm{backflow}}\to0^+$ and the positive-rate window vanishes, so the maximum transient dephasing becomes parametrically small ($\Gamma_{\max}\sim\overline{n}\chi^2/\kappa^2\to0$). 
In this limit, the cavity leaks too quickly to act as a memory and the dynamics is effectively Markovian. 

The bath-induced phase $\Phi(t)$ in Eq.~\eqref{eq:dephasing-coherence} is a coherent $\sigma_z$-rotation that we can mitigate against by working in the frame rotating at $\dot\Phi(t)$. 
This leaves pure dephasing, 
\begin{equation}
    \varrho(t)=\frac{1}{2}(\openone+e^{-\Gamma(t)} \sigma_x). 
\end{equation} 
Consider dynamics of the observable (coherence witness) $K=\sigma_x$ with $\varrho(0)=\ket+\,\bra+$,
\begin{equation} \label{eq:observable-expectation-lossy}
  \langle\sigma_x\rangle_t=e^{-\Gamma(t)}
    =\exp{-\overline{n}\,e^{-\kappa t}(1-\cos{2\chi t})},
\end{equation}
whose derivative
\begin{equation}
  \frac{d}{dt}\langle\sigma_x\rangle_t=-\gamma(t)\,e^{-\Gamma(t)}
\end{equation}
changes sign with $\gamma(t)$. 
In contrast to the $\kappa=0$ case, the revivals for $\kappa>0$ correspond to a failure of CP-divisibility and, in this model, to a revival of the coherence witness.
In fact, the coherence revives \textit{directly} through the dissipative channel, not through an asymmetry-conversion mechanism. 
Since $\varrho(t)$ commutes with $K=\sigma_x$, the asymmetry term vanishes identically and the motion is purely dissipative,  
\begin{equation}
     \mathcal{D}^\dagger_t[\sigma_x]=-\gamma(t)\sigma_x,
\end{equation} 
hence
\begin{equation}
  \langle \mathcal{D}^\dagger_t[\sigma_x]\rangle_t=-\gamma(t)\,e^{-\Gamma(t)}.
\end{equation}

The observable speed is
\begin{equation}
  v_K(t)=\frac{1}{2} \left| \tfrac{d}{dt}\langle\sigma_x\rangle_t \right|
    =\frac{1}{2}|\gamma(t)|\,e^{-\Gamma(t)},
\end{equation}
and the instantaneous bound is saturated at every instant,
\begin{equation}
  v_K(t)\le\frac{1}{2}\left(\|[\varrho,\sigma_x]\|_1
    +|\langle \mathcal{D}^\dagger_t[\sigma_x]\rangle_t|\right)
    =\frac{1}{2}|\gamma(t)|\,e^{-\Gamma(t)}.
\end{equation}

The dynamics~\eqref{eq:observable-expectation-lossy} are governed by $\Gamma(t)=\overline{n}\,e^{-\kappa t}(1-\cos{2\chi t})$, with two competing timescales. 
The dispersive oscillation $\cos{2\chi t}$ of period $\pi/\chi$ repeatedly drives the coherence, while the loss envelope $e^{-\kappa t}$ monotonically drains it on a timescale $1/\kappa$. 
The character of any einselection is set by their ratio $\chi/\kappa$, i.e., by how many oscillations survive before the envelope decays.

For strong dispersion, $\chi\gg\kappa$, many oscillation periods fit under a slowly-decaying envelope. 
Since $\Gamma$ is extremised where $\dot\Gamma=\gamma=0$, its global maximum is at the first zero of the rate,
\begin{equation}
  \Gamma_{\max}=\Gamma(t_{\mathrm{backflow}}),
\end{equation}
so, the strongest dephasing coincides with the onset of backflow.
For $\chi\gg\kappa$ this gives $\Gamma_{\max}\sim2\overline{n}$ (since $t_{\mathrm{backflow}} \sim\pi/2\chi$ makes the envelope $\sim1$ and $1-\cos{2\chi t_{\mathrm{backflow}}} \sim2$), so $\langle\sigma_x\rangle$ dips to $\sim e^{-2\overline{n}}$ before recohering. 

For heavy damping, $\kappa\gtrsim\chi$, the envelope collapses within the first oscillation period, leaving a single shallow dip followed by monotonic recoherence. 
The dephasing never builds up ($\Gamma_{\max}\sim\overline{n}\chi^2/\kappa^2\to0$ as $\kappa\to\infty$), and the dynamics is effectively Markovian.

The long-time behaviour of the intracavity conditional states is the same for both regimes.  
As $t\to\infty$, the mean intracavity photon number decays $\langle n(t)\rangle=\overline{n}\,e^{-\kappa t}\to0$, since $e^{-\kappa t}\to0$. 
Both conditional cavity states relax to the vacuum state, and their intracavity overlap returns to unity, so any distinguishability stored in the cavity itself is transient.

Within the description considered here, the leaky cavity drives the coherence witness toward its pointer value only transiently in the sense that as the cavity empties, the conditional cavity states become indistinguishable again.

\subsubsection{Independent boson model with Ohmic spectral density}
Thus far, we have realized pure dephasing within the Jaynes--Cummings model by taking its dispersive limit, with the environment provided by a single (Lorentzian) leaky cavity mode. 
We now replace this single mode by a continuum of bosonic modes, which allows the spectral density to be shaped, e.g., and set to be Ohmic. 
This is typically referred to as the independent boson model (solved in Mahan's book~\cite{Mahan2000} by canonical transformation) or the pure dephasing limit ($\Delta=0$) of the spin-boson model~\cite{Leggett1987}, whose dynamics are exactly solvable. 

Ultimately, the (pure-dephasing) Hamiltonian couples $\sigma_z$ to a continuum of bosonic modes~\cite{Palma1996, Unruh1995, Breuer2007},
\begin{equation}
  H = \tfrac{1}{2}\omega_q\sigma_z
    + \sum_k \omega_k b_k^\dagger b_k
    + \sigma_z\sum_k\left(g_k b_k + g_k^* b_k^\dagger\right).
\end{equation}
where $\omega_q$ is the qubit frequency, $b_k$ ($b_k^\dagger$) annihilates (creates) a boson in the $k$th bath mode of frequency $\omega_k$ (satisfying $[b_k,b_{k'}^\dagger]=\delta_{kk'}$), and $g_k$ is the coupling strength between the qubit and that mode. 
As in the dispersive case, the coupling is diagonal in $\sigma_z$, $[H,\sigma_z]=0$, so the populations in the $\{\ket{0},\ket{1}\}$ basis do not change over time and only the coherence $\varrho_{01}$ evolves.
Ref.~\cite{Breuer2007} works through this and derives the decoherence function \eqref{eq:bath-spectral-density}. 

As an example, we consider this result in the zero temperature limit of the Ohmic spectral density with an exponential ultraviolet cutoff at $\omega_c$, 
\begin{equation}
  J(\omega) = A\,\omega\,e^{-\omega/\omega_c},
\end{equation}
where $A$ is a dimensionless coupling that absorbs the overall normalisation constant. 
In this case, Eq.~\eqref{eq:bath-spectral-density} reduces to 
\begin{equation}
  \Gamma(t) = A\int_0^\infty \, d\omega\,
    \frac{e^{-\omega/\omega_c}}{\omega}\,\left(1-\cos{\omega t}\right).
\end{equation}
which evaluates to 
\begin{equation}
  \Gamma(t) = \frac{A}{2}\,\ln\,\left(1+\omega_c^2 t^2\right),
\end{equation}
with the two limiting regimes
\begin{align}
  \Gamma(t) &\approx \tfrac{A}{2}\,\omega_c^2 t^2, & \omega_c t &\ll 1,\\
  \Gamma(t) &\approx A\,\ln(\omega_c t), & \omega_c t &\gg 1, 
\end{align}
i.e., a quadratic short-time onset and long-time unbounded logarithmic growth. 

The coherence behaves as 
\begin{align}
  |\varrho_{01}(t)| &= |\varrho_{01}(0)\,e^{-\Gamma(t)}| \\ 
                 &= |\varrho_{01}(0)|\,\left(1+\omega_c^2 t^2\right)^{-A/2}
  \xrightarrow{\,t\to\infty\,} 0,
\end{align}
i.e., a power-law decay $\sim t^{-A}$ at long times ($\omega_c t\gg1$).
In contrast to the Lorentzian single-mode case, where $\Gamma(t)$ oscillates and returns to zero (transient dephasing and full recoherence), the Ohmic continuum gives monotonic, permanent decoherence. 
Einselection is therefore permanent, and the witness $\la\sigma_x\ra=e^{-\Gamma(t)}$ relaxes monotonically to its pointer value $0$.

The canonical decoherence rate
\begin{equation}
  \gamma(t) = \dot\Gamma(t) = \frac{A\,\omega_c^2 t}{1+\omega_c^2 t^2} > 0
  \qquad (t>0)
\end{equation}
is strictly positive for all $t$, so the map is CP-divisible although it produces permanent decoherence. 
From this, we see that Markovian versus non-Markovian and permanent versus transient einselection are thus independent properties, controlled respectively by the sign structure of $\gamma(t)$ and by the shape of $J(\omega)$ that governs the long-time limit of $\Gamma(t)$. 

\subsubsection{Structured continuum}
This inspires us to combine the two behaviors and construct the spectral density given by Eq.~\eqref{eq:structured-J} in the main text. 

At $T=0$, the Ohmic low-frequency contribution supplies the (logarithmic) divergence $\Gamma(t) \to \infty$ via vacuum fluctuations, and the Lorentzian resonance at $\omega_0$ supplies the nonmonotonicity of the rate $\gamma$ with revivals. 

While we note that the underdamped Brownian oscillator~\cite{Leggett1987, Garg1985JCP} density combines both terms, 
\begin{equation}
    J(\omega) = \frac{C\omega}{(\omega^2-\omega_0^2)^2+\lambda^2\omega^2}, 
\end{equation}
we have chosen to retain the explicit additive structure of Eq.~\eqref{eq:structured-J} in the main text to emphasise the two conditions for permanent einselection. 

\section{Non-Markovian enhancement and delay bounds}
\label{app:nonmarkovian-bounds}

\subsection{Direct enhancement}
In Section~\ref{sec:non-markovian-modification-direct-enhancement}, we extended the time-homogeneous amplitude-damping dynamics considered in Sec.~\ref{sec:instantaneous-bound} (whose details are given in Appendix~\ref{app:amplitude-damping}) to a time-dependent canonical decoherence rate. The time-local master equation~\eqref{eq:amplitude-damping} (reproduced as Eq.~\eqref{eq:app-amplitude-damping}) is modified only to have a time-dependent canonical decoherence rate $\gamma(t)$, 
\begin{align}
    \dot{\varrho}(t) &= - i[\sigma_z,\varrho(t)]
    + \gamma(t) \left( \sigma_-\varrho(t)\sigma_+ - \frac{1}{2} \left\{ \sigma_+\sigma_-, \varrho(t) \right\} \right),
\end{align}
with $H=\sigma_z$, $L=\sigma_-$, $K=\sigma_z$, $\varrho(0)=\ket{1}\!\bra{1}$. 
Analogous to the time-independent dynamics detailed in Appendix~\ref{app:amplitude-damping}, the state remains diagonal throughout the evolution, but now its excited-state population
\begin{equation}
    \varrho_{11}(t) = e^{-\Gamma(t)}
\end{equation}
is determined by the integrated canonical decoherence rate
\begin{align}
    \Gamma(t) &=  \int_0^t ds\,\gamma(s). 
\end{align}
Consequently, the expectation value of $(K=\sigma_z$ is
\begin{align}
    \la K\ra_t &= 1-2e^{-\Gamma(t)},
\end{align} 
with an instantaneous speed~\eqref{eq:speed-definition}, 
\begin{align}
    v_K(t) &= \left|\gamma(t)\right|e^{-\Gamma(t)}.
\end{align}

As in Appendix~\ref{app:amplitude-damping}, since $\varrho(t)$ and $K$ remain diagonal in the same basis, $[\varrho(t),K]=0$, and the coherent current vanishes, so the observable motion is therefore generated entirely by the dissipative current. 
As such, the instantaneous observable-speed bound~\eqref{eq:instantaneous-bound} is again saturated,
\begin{align}
    v_K(t) = \frac{1}{2} \left|\la\mathcal{D}_t^\dagger[K]\ra_t\right|.
\end{align}

As a simple example, choose the time-dependent canonical decoherence rate 
\begin{align}
    \gamma_{\mathrm{NM}}(t) = \gamma_0\sin{(\Omega t)}, \qquad \gamma_0>0, 
\end{align}
so the rate becomes negative on alternating half-periods, and the evolution is therefore CP-indivisible on those intervals. 
Nevertheless, the corresponding integrated decoherence rate satisfies
\begin{align}
    \Gamma_{\mathrm{NM}}(t) = \frac{\gamma_0}{\Omega} \left[ 1-\cos{(\Omega t)} \right] \geq 0,  
\end{align}
so that $0<e^{-\Gamma_{\mathrm{NM}}(t)}\leq1$, and the dynamical map from the initial time to $t$ remains CPTP despite the breakdown of CP-divisibility.

During each negative-rate interval, the excited-state population increases and the signed motion of $\la\sigma_z\ra_t$ reverses.
However, this reversal alone does not establish an enhancement of the absolute observable speed. 

Compare these dynamics with a matched CP-divisible reference defined by Eq.~\eqref{eq:matched-cp-rate}. 
This reference dynamics is identical, except for the sign of the canonical decoherence rate, $\gamma_{\mathrm{M}}(t)\geq 0$, so that the reference evolution is CP-divisible.

The corresponding integrated canonical decoherence rate, excited-state population, and observable speed of the matched CP-divisible dynamics are 
\begin{align}
    \Gamma_{\mathrm{M}}(t) &= \int_0^t ds\,\left|\gamma_{\mathrm{NM}}(s)\right|,  \\
    \varrho_{11}^{\mathrm{M}}(t) &= e^{-\Gamma_{\mathrm{M}}(t)}, \\
    v_K^{\mathrm{M}}(t) &= \left|\gamma_{\mathrm{NM}}(t)\right| e^{-\Gamma_{\mathrm{M}}(t)}.
\end{align}
For the accumulated canonical-rate negativity defined in Eq.~\eqref{eq:accumulated-negativity}, Eqs.~\eqref{eq:Gamma-difference}, \eqref{eq:nm-direct-enhancement-rho11}, and \eqref{eq:nm-direct-enhancement-speed} hold. 
Moreover, since $F(0,t)\geq0$, the CP-indivisible dynamics satisfies
\begin{align}
    v_K^{\mathrm{NM}}(t) \geq v_K^{\mathrm{M}}(t)
    \label{eq:nm-speed-bound}
\end{align}
for $t\geq 0$. 
The inequality is strict whenever canonical-rate negativity has
previously accumulated, $F(0,t)>0$, and the instantaneous rate magnitude is nonzero. 

\begin{figure}
    \centering
    \includegraphics[trim=19cm 1cm 19cm 2cm, clip=true, width=\columnwidth]{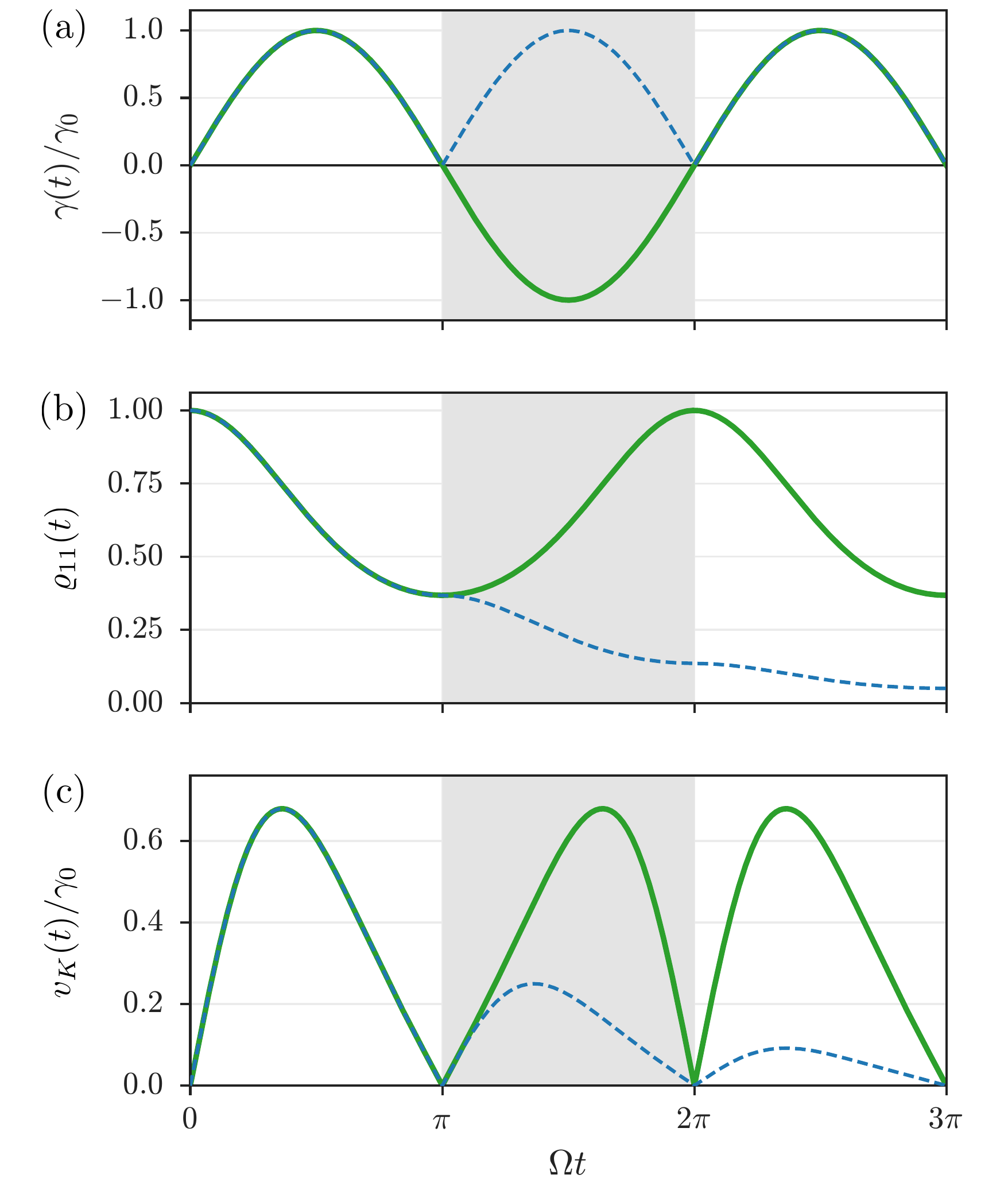}
    \caption{Matched CP-indivisible and CP-divisible amplitude-damping dynamics. 
    Solid green curves show the CP-indivisible evolution generated by $\gamma_{\mathrm{NM}}(t)=\gamma_0\sin{(\Omega t)}$, while dotted blue curves show the matched CP-divisible reference with $\gamma_{\mathrm{M}}(t)=|\gamma_{\mathrm{NM}}(t)|$. 
    The shaded region indicates the interval for which $\gamma_{\mathrm{NM}}(t)<0$. 
    (a) Canonical decoherence rates normalized by $\gamma_0$. 
    The two rates coincide outside the shaded interval. 
    (b) Excited-state population. 
    The CP-indivisible dynamics exhibits a population revival, in contrast to the monotonic decay of the CP-divisible reference. 
    (c) Observable speed for $K=\sigma_z$, normalized by $\gamma_0$. 
    The enhanced speed under CP-indivisible dynamics is due to the larger population retained or restored through the accumulated negative canonical decoherence rate, rather than from an increase in the instantaneous magnitude of the decoherence rate.
    }
    \label{fig:nonmarkovian-speedup}
\end{figure}

Figure~\ref{fig:nonmarkovian-speedup} illustrates this comparison for $\gamma_0/\Omega=1/2$. 
Before the first negative-rate interval, the two dynamics coincide. During the negative-rate interval, the CP-indivisible evolution restores the excited-state population, whereas the population of the CP-divisible reference continues to decrease.
The resulting population difference persists after the canonical decoherence rate becomes positive again and produces a correspondingly larger observable speed.

The enhancement in Eq.~\eqref{eq:nm-speed-bound} does not arise from a larger instantaneous decoherence-rate magnitude, which is identical for the two dynamics. 
Instead, the accumulated negative canonical decoherence rate retains or restores the excited-state population, leaving a larger population available for dissipative motion. 
In this model, CP-indivisibility can therefore act as a resource for the instantaneous observable speed.

This result is model and observable-dependent. A negative canonical decoherence rate reverses a signed dissipative current, but it need not increase the absolute speed of an arbitrary observable when coherent motion or additional dissipative channels are present. Moreover, instantaneous speed enhancement does not necessarily produce a tighter finite-time observable quantum speed limit. 
At $\Omega t=2\pi$, for example, the CP-indivisible solution returns to its initial population, so the net observable displacement vanishes despite nonzero observable motion at intermediate times.

\subsection{Indirect enhancement}
In Section~\ref{sec:non-markovian-modification-indirect-enhancement}, while the environment does not directly drive the pointer observable $K_{\mathrm{ptr}}=\sigma_z$, it can indirectly enhance the observable speed. 
For a decoherence operator $L\propto\sigma_z$, one has $\mathcal{D}_t^\dagger[K_{\mathrm{ptr}}]=0$, so the dissipative current vanishes identically and the instantaneous bound~\eqref{eq:instantaneous-bound} reduces to the coherent term alone,
\begin{equation}
    v_{K_{\mathrm{ptr}}}(t)
    \leq
    \tfrac{1}{2}\left\|[\varrho(t),K_{\mathrm{ptr}}]\right\|_1 .
    \label{eq:pointer-asymmetry-bound}
\end{equation}
For a qubit the right-hand side is the $\ell_1$ coherence in the eigenbasis of $K_{\mathrm{ptr}}$,
\begin{equation}
    \tfrac{1}{2}\left\|[\varrho(t),K_{\mathrm{ptr}}]\right\|_1
    = C_{\ell_1}^{(\mathrm{ptr})}[\varrho(t)] .
    \label{eq:pointer-coherence-asymmetry}
\end{equation}
The dephasing environment thus leaves the pointer observable unchanged while modifying the coherence that bounds its subsequent coherent motion. This is the indirect mechanism by which non-Markovianity can enhance observable speed of Sec.~\ref{sec:non-markovian-modification}, as exemplified in the following prepare-and-readout protocol, which is similar in spirit to the setup in Ref.~\cite{Galve2016SRep}.

\subsubsection{Preparation step}

To isolate this effect, consider a preparation with $H_p=\sigma_z$ and $\varrho(0)=\ket{+}\!\bra{+}$. 
Since $[H_p,K_{\mathrm{ptr}}]=0$ and $\mathcal{D}_t^\dagger[K_{\mathrm{ptr}}]=0$, $v_{K_{\mathrm{ptr}}}(t)=0$ and the pointer observable is stationary while its pointer-basis coherence evolves according to
\begin{equation} \label{eq:app-dephasing-coherence}
    C_{\ell_1}^{(\mathrm{ptr})}(\varrho(t)) = e^{-\Gamma(0,t)},
    \qquad
    \Gamma(0,t)=\int_0^t ds\,\gamma(s).
\end{equation}
Whenever $\gamma(t)<0$, the pointer-basis coherence increases. 
Negative canonical decoherence rates therefore restore the asymmetry available for coherent motion without producing any direct motion of $\la K_{\mathrm{ptr}}\ra_t$. 
The coherence revivals found in Sec.~\ref{sec:non-markovian-modification} therefore represent an available resource for coherent motion rather than motion of the pointer observable itself. 
Compared with the matched CP-divisible reference $\gamma_{\mathrm{M}}(t)=|\gamma_{\mathrm{NM}}(t)|$ \eqref{eq:matched-cp-rate}, the integrated rates differ by $\Gamma_{\mathrm{M}}(0,t)-\Gamma_{\mathrm{NM}}(0,t)=2F(0,t)$ \eqref{eq:Gamma-difference}, and hence
\begin{equation}
    C_{\ell_1}^{\mathrm{NM}}(\varrho(t))
    = e^{2F(0,t)}\,C_{\ell_1}^{\mathrm{M}}(\varrho(t)) ,
    \label{eq:coherence-enhancement}
\end{equation}
so the accumulated negativity quantifies the coherence retained by the CP-indivisible evolution relative to its matched CP-divisible reference.

\subsubsection{Cyclic rate and comparison}
Take the sinusoidal canonical decoherence rate
\begin{equation}
    \gamma_{\mathrm{NM}}(t)=\gamma_0\sin{(\Omega t)},
    \label{eq:sinusoidal-rate}
\end{equation}
and evaluate the dynamics at the times $t_n=2\pi n/\Omega$, $n\in\mathbb{N}$, which complete an integer number of rate cycles. 
Then $\Gamma_{\mathrm{NM}}(0,t_n)=0$ while $\Gamma_{\mathrm{M}}(0,t_n)=\int_0^{t_n} ds\,|\gamma_0\sin{\Omega s}| = 4n\gamma_0/\Omega$, so
\begin{equation}
    C_{\ell_1}^{\mathrm{NM}}(\varrho(t_n))=1,
    \qquad
    C_{\ell_1}^{\mathrm{M}}(\varrho(t_n))=e^{-4n\gamma_0/\Omega},
    \label{eq:cyclic-coherence}
\end{equation}
and the accumulated negativity is $F(0,t_n)=\tfrac12\Gamma_{\mathrm{M}}(0,t_n) =2n\gamma_0/\Omega$. 
The CP-indivisible evolution therefore returns to unit pointer-basis coherence after every complete cycle, whereas the reference coherence decreases exponentially in the number of cycles.

\subsubsection{Readout step}
The times $t_n$ are those at which the canonical decoherence rate vanishes, $\gamma_{\mathrm{NM}}(t_n)=\gamma_{\mathrm{M}}(t_n)=0$. 
Terminating the dephasing preparation there and applying a normalized unitary drive to the prepared state $\varrho(t_n)$ isolates the conversion of restored coherence into observable speed, with no dissipative contribution at the switching instant. 
I.e., the readout step is unitary and hence CPTP. 
The described prepare-and-readout protocol exhibits the restored asymmetry as available observable speed at the coherent bound.

During preparation the drift $H_p=\sigma_z$ precesses the Bloch vector about the $z$ axis at angular frequency $2$, while the dephasing keeps it in the equatorial plane. 
Starting from $\varrho(0)=\ket{+}\!\bra{+}$ (azimuth $\phi=0$), the prepared state at the comparison time is
\begin{align}
    \varrho(t_n)&=\tfrac12\left(\openone+\mathbf r_n\cdot\bm\sigma\right), \\
    \mathbf r_n &= C_{\ell_1}^{(\mathrm{ptr})}(\varrho(t_n))\,
    \left(\cos{\phi_n},\,\sin{\phi_n},\,0\right),
    \label{eq:prepared-bloch}
\end{align}
with $\phi_n=2t_n$ and equatorial radius equal to the pointer-basis coherence $C_{\ell_1}^{(\mathrm{ptr})}(\varrho(t_n))$.

The coherent bound~\eqref{eq:pointer-asymmetry-bound} is saturated when the drive is aligned with the sign operator of $-i[\varrho(t_n),K_{\mathrm{ptr}}]$. 
For the equatorial state~\eqref{eq:prepared-bloch} this sign operator is $\hat S_n=\sin{\phi_n}\,\sigma_x-\cos{\phi_n}\,\sigma_y$, so the choice
\begin{align}
    H_c(t_n) &=\sin{\phi_n}\,\sigma_x-\cos{\phi_n}\,\sigma_y \\
    &=\sin{(2t_n)}\,\sigma_x-\cos{(2t_n)}\,\sigma_y,
    \label{eq:readout-hamiltonian}
\end{align}
with $\norm{H_c(t_n)}_\infty=1$ saturates the Schatten-H\"{o}lder inequality. 
This orientation drives the Bloch vector along the $z$ axis, i.e., it generates motion of the pointer observable $\la K_{\mathrm{ptr}}\ra_t$ at the maximal coherent rate. 
The restored equatorial coherence is thereby converted directly into pointer-observable speed. 
With this drive,
\begin{equation}
    v_{K_{\mathrm{ptr}}}(t_n)
    = C_{\ell_1}^{(\mathrm{ptr})}(\varrho(t_n))
    = \tfrac12\left\|[\varrho(t_n),K_{\mathrm{ptr}}]\right\|_1 ,
\end{equation}
so the coherent bound is attained at each comparison time. For the matched pair,
\begin{equation}
    v_{K_{\mathrm{ptr}}}^{\mathrm{NM}}(t_n)=1,
    \qquad
    v_{K_{\mathrm{ptr}}}^{\mathrm{M}}(t_n)=e^{-4n\gamma_0/\Omega},
    \label{eq:coherence-speed-enhancement}
\end{equation}
and, using $F(0,t_n)=2n\gamma_0/\Omega$,
\begin{equation}
    v_{K_{\mathrm{ptr}}}^{\mathrm{NM}}(t_n)
    = e^{2F(0,t_n)}\,v_{K_{\mathrm{ptr}}}^{\mathrm{M}}(t_n)
    \;\geq\; v_{K_{\mathrm{ptr}}}^{\mathrm{M}}(t_n),
    \label{eq:coherence-to-speed}
\end{equation}
which realizes the indirect enhancement $e^{2F(0,t)}$ of the main text through the coherent current.

\subsection{Derivation of the delay bound}
\label{app:delay-bound}
Here, we derive the lower bound~\eqref{eq:delay-bound} on the delay of a relaxing quantity. 
The setting is a single canonical channel with a matched pair~\eqref{eq:matched-cp-rate} of dynamics, $\gamma_{\mathrm{M}}(t) = |\gamma_{\mathrm{NM}}(t)|$, and a relaxing quantity that decays as in Eq.~\eqref{eq:relaxing-quantity}, $\mathcal{R}_i(t) = \mathcal{R}(0)\,e^{-\Gamma_i(0,t)}$ with $\Gamma_i(0,t)=\int_0^t ds\,\gamma_i(s)$ for $i\in\{\mathrm{M},\mathrm{NM}\}$. 
We assume permanent einselection, $\Gamma_i(0,t)\to\infty$, so that the threshold is reached at a finite time for every $\epsilon$.

Writing $r \equiv \ln[\mathcal{R}(0)/\epsilon]$, the times at which the threshold is crossed are the smallest solutions of 
\begin{equation}
    \Gamma_{\mathrm{M}}(0,t^{\mathrm{M}}_\epsilon) = r,
    \qquad
    \Gamma_{\mathrm{NM}}(0,t^{\mathrm{NM}}_\epsilon) = r.
    \label{eq:app-threshold-crossing}
\end{equation}
From the matched-rate condition,
\begin{align}
    \Gamma_{\mathrm{M}}(0,t) - \Gamma_{\mathrm{NM}}(0,t)
    &= \int_0^t ds\, \left(|\gamma_{\mathrm{NM}}(s)| - \gamma_{\mathrm{NM}}(s)\right) \nonumber \\
    &= 2F(0,t) \geq 0,
    \label{eq:app-deficit}
\end{align}
so $\Gamma_{\mathrm{NM}}(0,t)\leq\Gamma_{\mathrm{M}}(0,t)$ for all $t$. Evaluating at $t^{\mathrm{M}}_\epsilon$ and using Eq.~\eqref{eq:app-threshold-crossing},
\begin{equation}
    \Gamma_{\mathrm{NM}}(0,t^{\mathrm{M}}_\epsilon)
    = r - 2F(0,t^{\mathrm{M}}_\epsilon) \leq r,
\end{equation}
so the non-Markovian solution has not yet reached the threshold at $t^{\mathrm{M}}_\epsilon$, giving $t^{\mathrm{NM}}_\epsilon \geq t^{\mathrm{M}}_\epsilon$ and hence $\Delta t_\epsilon\geq 0$.

The exponent that remains to be accumulated between the two threshold-crossings is fixed by Eqs.~\eqref{eq:app-threshold-crossing} and~\eqref{eq:app-deficit},
\begin{align}
    \Gamma_{\mathrm{NM}}(0,t^{\mathrm{NM}}_\epsilon)
    - \Gamma_{\mathrm{NM}}(0,t^{\mathrm{M}}_\epsilon)
    &= r - \left[r - 2F(0,t^{\mathrm{M}}_\epsilon)\right] \nonumber \\
    &= 2F(0,t^{\mathrm{M}}_\epsilon).
    \label{eq:app-residual}
\end{align}
Writing the left-hand side as a time integral of the canonical decoherence rate and bounding the integrand by its supremum over the catch-up window,
\begin{equation}
    2F(0,t^{\mathrm{M}}_\epsilon)
    = \int_{t^{\mathrm{M}}_\epsilon}^{t^{\mathrm{NM}}_\epsilon}
        ds\,\gamma_{\mathrm{NM}}(s)
    \;\leq\;
    \gamma_{\max}\,\left(t^{\mathrm{NM}}_\epsilon - t^{\mathrm{M}}_\epsilon\right),
    \label{eq:app-integral-bound}
\end{equation}
with 
\begin{equation}
    \gamma_{\max} \equiv
    \sup_{s\in[t^{\mathrm{M}}_\epsilon,\,t^{\mathrm{NM}}_\epsilon]}
        \gamma_{\mathrm{NM}}(s),
\end{equation}
where the inequality uses only $\gamma_{\mathrm{NM}}(s)\leq\gamma_{\max}$ pointwise and holds irrespective of the sign of $\gamma_{\mathrm{NM}}$ within the window. 
Rearranging Eq.~\eqref{eq:app-integral-bound} gives the bound of the main text,
\begin{equation}
    \Delta t_\epsilon
    = t^{\mathrm{NM}}_\epsilon - t^{\mathrm{M}}_\epsilon
    \;\geq\;
    \frac{2F(0,t^{\mathrm{M}}_\epsilon)}{\gamma_{\max}}.
    \label{eq:app-delay-bound}
\end{equation}
Equality holds when $\gamma_{\mathrm{NM}}(s)=\gamma_{\max}$ throughout the catch-up window $[t^{\mathrm{M}}_\epsilon,t^{\mathrm{NM}}_\epsilon]$. 
Replacing $\gamma_{\max}$ by the global supremum of $\gamma_{\mathrm{NM}}$ over $[0,t^{\mathrm{NM}}_\epsilon]$ yields a weaker bound whose denominator no longer depends on the catch-up window.

The local supremum gives the sharper estimate. 
If the CP-indivisibility is confined to $t\leq t^{\mathrm{M}}_\epsilon$, then $F(0,t^{\mathrm{M}}_\epsilon)=F_\infty$ is the total accumulated negativity, 
$F_\infty = \lim_{t\to \infty} F(0,t) = \int_0^\infty ds\, [-\gamma_\mathrm{NM}(s)]_+ $. 
If the relaxation rate has also reached (and does not exceed) its asymptotic value $\gamma_\infty$ beyond $t^{\mathrm{M}}_\epsilon$, then $\gamma_{\max}=\gamma_\infty$ and the bound reduces to $\Delta t_\epsilon \geq 2F_\infty/\gamma_\infty$, independent of $\epsilon$.

\bibliography{qsl_references}

\end{document}